\documentclass[twocolumn,showpacs,preprintnumbers,amsmath,amssymb,superscriptaddress]{revtex4-1}
\usepackage{amssymb,bm}
\usepackage{booktabs}
\usepackage{graphicx,color}
\begin{document}

\title{Relics of Minijets amid Anisotropic Flows in High-energy Heavy-ion Collisions}

\author{Longgang Pang}

\affiliation{Institute of Particle Physics and Key Laboratory of Quarks and Lepton Physics (MOE), Central China Normal University, Wuhan 430079, China}
\affiliation{Nuclear Science Division, MS 70R0319, Lawrence Berkeley National Laboratory, Berkeley, CA 94720}

\author{Qun Wang}
\affiliation{Interdisciplinary Center for Theoretical Study and Department of Modern Physics, University of Science and Technology of China, Hefei 230026, China}

\author{Xin-Nian Wang}
\affiliation{Institute of Particle Physics and Key Laboratory of Quarks and Lepton Physics (MOE), Central China Normal University, Wuhan 430079, China}
\affiliation{Nuclear Science Division, MS 70R0319, Lawrence Berkeley National Laboratory, Berkeley, CA 94720}

\begin{abstract}

	Two dimensional low-$p_T$ dihadron correlations in azimuthal angle $\phi$ and pseudo-rapidity $\eta$ in high-energy
	heavy-ion collisions are investigated	within both the HIJING Monte Carlo model and an event-by-event (3+1)D ideal
	hydrodynamic model. Without final-state interaction and collective expansion, dihadron correlations
	from HIJING simulations have a typical structure from minijets that contains a near-side two-dimensional peak 
	and an away-side ridge along the $\eta$-direction. In contrast, event-by-event (3+1)D ideal
	hydrodynamic simulations with fluctuating initial conditions from the HIJING+AMPT model
	produce a strong dihadron correlation that has an away-side as well as a near-side ridge. Relics of
	intrinsic dihadron correlation from minijets in the initial conditions still remain as superimposed on the two ridges. 
	By varying initial conditions from HIJING+AMPT, we study effects of minijets, non-vanishing
	initial flow and longitudinal fluctuation on the final state dihadron correlations. With a large rapidity gap, one can 
	exclude near-side correlations from minijet relics and dihadron correlations 
	can be described by the superposition of harmonic flows up to the 6th order.	 When long-range correlations with a large rapidity gap 
	are subtracted from short-range correlations with a small rapidity gap, the remaining near-side dihadron correlations result solely 
	from relics of minijets. Low transverse momentum hadron yields per trigger ( $p_T^{\rm trig} <4$ GeV/$c$, 
	$p_T^{\rm asso}<2$ GeV/$c$) in central heavy-ion collisions are significantly enhanced over that in p+p collisions 
	while widths in azimuthal angle remain the same, in qualitative agreement with experimental data.



\end{abstract}

\maketitle

\section{Introduction}

Jet production has been the focus of many studies in high-energy heavy-ion collisions. Minijets with moderate transverse momentum contribute to an increasingly large fraction of the initial energy density of partons \cite{Blaizot:1987nc,Kajantie:1987pd,Eskola:1988yh,Wang:1990qp,Eskola:1995bp,Wang:2000bf}  that thermalize (or partially thermalize) and form a new state of matter called quark-gluon plasma (QGP). High transverse momentum jets, on the other hand, have been proposed as hard probes of the dense matter formed in high-energy heavy-ion collisions through jet quenching \cite{Wang:1991xy}. Propagation of the energy-momentum deposited in the medium by jet-medium interaction can also induce medium excitations which can be used to study transport properties of the QGP \cite{CasalderreySolana:2004qm,Stoecker:2004qu,Chaudhuri:2005vc,Neufeld:2008fi,
Betz:2008js,Neufeld:2008dx,Neufeld:2009ep,Qin:2009uh,Betz:2008ka,Li:2010ts,Ma:2010dv}. 
Measurements of dihadron correlations have been proposed in all of jet-related studies in high-energy heavy-ion collisions, 
from anisotropic expansion of hot spots of minijets to jet quenching and jet-induced medium excitations.

Within the lowest-order perturbative QCD (pQCD) parton model, jets in hadronic and nuclear collisions are produced in pairs. They are approximately back-to-back in the azimuthal angle $\Delta \phi\approx \pi$. For moderate transverse momentum far away from the kinetic bounds, the rapidity differences between the two jets can vary in a large region $\Delta \eta < 2 \ln (\sqrt{s}/2p_T)$ due to different initial momentum fractions of the two colliding partons. The corresponding two-dimensional dihadron correlations in azimuthal angle $\phi$ and pseudo-rapidity $\eta$ from dijets therefore have a typical structure intrinsic to dijet events. It has one near-side peak at $(\Delta\phi=0, \Delta\eta=0$) by hadron pairs from the same jet and one away-side ridge along the $\Delta\eta$ direction at $\Delta\phi = \pi$ by hadron pairs from two jets separately. The shapes of the near-side peak and away-side ridge in $\Delta\phi$ are approximately Gaussian because of initial state radiation, intrinsic transverse momentum of the initial partons and final state jet fragmentation. 

The mechanism for initial jet production in high-energy heavy-ion collisions remains basically the same as in nucleon-nucleon collisions except some cold nuclear modification of the initial parton distributions. The initially produced jet partons, however, will have to traverse a strongly interacting matter that is formed in heavy-ion collisions. Minijets with small and moderate transverse momentum will interact with each other and other soft partons from multiple coherent nucleon-nucleon collisions. They become part of the bulk QGP medium approaching {\it local} thermal equilibrium. Relics of the intrinsic parton correlations from the initial minijets should appear in the final-state dihadron correlations from the bulk medium that does not reach complete {\it global} thermalization. Jets with large transverse momentum, on the other hand, will also interaction with the bulk medium as they propagate through the medium. Such multiple interactions of high $p_T$ jet partons will lead to parton energy loss and suppression of both leading hadrons \cite{Adcox:2001jp,Adler:2003qi,Adams:2003kv} and the back-to-back dihadron correlation \cite{Adler:2002tq}, a phenomenon known as jet quenching. Energy lost by high $p_T$ jets, furthermore, is transferred to soft partons via radiated gluons and recoil medium partons. These soft partons in the form of jet-induced medium excitation on top of the expanding fireball should also affect the final dihadron correlations in heavy-ion collisions.

Dihadron correlations in high-energy heavy-ion collisions were measured at the Relativistic Heavy-ion Collisions (RHIC) in search for jet-induced medium excitations. After subtraction of contributions from elliptic flow due to collective expansion of an anisotropic fireball, a double-bump structure was found in the away-side low-$p_T$ dihadron correlations \cite{Adams:2005ph,Ulery2006581,Adler:2005ee,Adare:2008ae,Abelev:2008ac,TakahitoTodorokiforthePHENIX:2013kia} around $\Delta\phi = \pi \pm 1.1$ in Au+Au collisions at RHIC. In the meantime, a near-side ridge structure along the $\eta$ direction in the two-dimensional dihadron correlation is also observed \cite{Agakishiev:2011pe,Putschke:2007mi,Abelev:2009af,Alver:2009id} underneath a peak from the residual correlation of jets that have survived jet quenching and thermalization. Many theoretical explanations have been proposed to explain the near-side ridge \cite{Armesto:2004pt,Chiu:2005ad,Shuryak:2007fu,Pantuev:2007sh,Gavin:2008ev,Wong:2011qr} and away-side double-bump structure \cite{CasalderreySolana:2004qm,Stoecker:2004qu,Li:2010ts,Ma:2010dv,Koch:2005sx,Chaudhuri:2005vc,Antinori:2005tu} in heavy-ion collisions. The consensus now is that both the near-side ridge and away-side double-bump have a common origin. They arise mainly from higher-order harmonic flows, especially the triangle flow \cite{Alver:2010gr} due to anisotropic expansion of the initial energy density with geometric fluctuations. This picture of  expansion of anisotropic fireballs is unambiguously demonstrated by measurements of dihadron correlations and harmonic flows in Pb+Pb collisions at the Large Hadron Collider (LHC) \cite{Aamodt:2011by,Chatrchyan:2012wg,Abelev:2012hxa,ATLAS:2012at,Chatrchyan:2011eka}. With large rapidity gap, one can exclude contributions from relics of minijets in the dihadron correlations, at least in the near-side. The measured dihadron correlations are identical to that from superposition of harmonic flows up to the sixth order. The same picture also emerges from AMPT Monte Carlo simulations \cite{Ma:2010dv,Xu:2011jm}. However, how minijets survive the initial thermalization and hydrodynamic expansion and contribute to dihadron correlations in final states still remains unexplored. Though the fluctuation of initial energy density in the transverse direction is understood to lead to the final anisotropic flows, how fluctuations in the longitudinal direction (pseudo-rapidity) influence the shape of the ridge structure in rapidity is still not clear. These are the focus our investigations in this paper.

We will first use HIJING Monte Carlo model \cite{Wang:1991hta,Gyulassy:1994ew,Deng:2010mv,Xu:2012au} to study the intrinsic dihadron correlations in p+p and A+A collisions from minijets without final state interaction. We then use a recently developed (3+1)D hydrodynamic model \cite{Pang:2012he} to investigate how locally thermalized minijets in an expanding anisotropic fireball contribute to the final dihadron correlations  in heavy-ion collisions. In order to study the contributions of anisotropic flows to di-hadron correlations, we will use the HIJING model plus parton cascade in the AMPT model \cite{Zhang:1999bd} to provide fluctuating initial conditions for event-by-event hydrodynamic simulations \cite{Pang:2012he}. Fluctuating initial conditions with the MC-Glauber model \cite{Miller:2007ri}, MC-KLN model \cite{Drescher:2006ca}, NeXus model \cite{Drescher:2000ec,Hama:2004rr},  EPOS model \cite{Werner:2010aa} and UrQMD model \cite{Bass:1998ca} have been used for the study of two-dimensional dihadron correlations \cite{Werner:2012xh,Bozek:2012en,Qian:2012qn} and anisotropic flows \cite{Petersen:2010cw,Qiu:2011hf,Schenke:2011bn} in heavy-ion collisions with event-by-event hydrodynamic simulations. In this paper, we will focus on the relics of minijets and the effect of longitudinal fluctuations on the final low-$p_T$ dihadron correlations in heavy-ion collisions using an ideal (3+1)D hydrodynamics with fluctuating initial conditions from the HIJING+AMPT model. Introduction of viscosity into (3+1)D event-by-event hydrodynamic simulations can change the numerical results quantitatively. However, our conclusions will remain qualitatively the same.

The remainder of this paper is organized as follows. In Section II, we use HIJING model to study the structure of two-dimensional dihadron correlations in p+p and Au+Au collisions at RHIC without hydrodynamic expansion. We also investigate the influence of jet quenching on the modification of dihadron correlations from minijets. We then calculate dihadron correlations within an ideal (3+1)D event-by-event hydrodynamic model with fluctuating initial conditions as given by HIJING+AMPT model in Section III. To isolate contributions to dihadron correlations from minijets especially on the near side, we subtract dihadron correlation with large rapidity gap from that with small rapidity gap. The obtained charged particle yields per trigger from ideal (3+1)D event-by-event hydrodynamic simulations are compared to PHENIX (Au+Au $\sqrt{s}=200$ GeV/n) and CMS (Pb+Pb $\sqrt{s}=2.76$ TeV/n) data. In Secction IV, We will study the influence of minijets and their fluctuations in the longitudinal direction on the dihadron correlations by varying the fluctuation in both initial energy density and flow velocity. We also compare direct dihadron correlations with that reconstructed from harmonic flows. Finally we give our summary and conclusions in Sec. V.

\section{Dihadron correlations in p+p and A+A collisions from HIJING}

Jet production with large transverse momentum transfer can be described by perturbative QCD (pQCD). At the lowest order of pQCD, jets in hadronic and nuclear collisions are normally produced in pairs. They are back-to-back in azimuthal angle $\phi$ with the differential cross section as given by \cite{Eichten:1984eu}
\begin{equation}
\frac{d\sigma_{\rm jet}}{dp_T^2dy_1dy_2}=\sum _{a,b} x_1 f_a(x_1,p_T^2) x_2 f(x_2,p_T^2) d\sigma_{ab}/d\hat t,
\end{equation}
where $\sigma_{ab}$ are the pQCD cross section of two-parton scatterings. The rapidities $y_{1,2}$ of the two jets are determined by the longitudinal momentum fractions $x_{1,2}$ of the two colliding partons,
\begin{equation}
x_{1,2}=\frac{2p_T}{\sqrt{s}}\left(e^{\pm y_1}+e^{\pm y_2}\right),
\end{equation}
and the final jet transverse momentum $p_T$.
For jet production with moderate transverse momentum far from the kinetic limit $p_T\ll \sqrt{s}/2$ or $x_{1,2}\ll 1$, parton distribution functions have a power-law behavior $f_a(x,p_T^2) \sim 1/x^{1+\alpha}$ with $\alpha > 0$ \cite{Gluck:1994uf} and the parton scatterings are dominated by $t$ or $u$-channels. In this kinetic region, the rapidity distribution of jets will have an approximate plateau with a half-width $\sim \ln\sqrt{s}/2p_T$. In this lowest order pQCD collinear parton model, dijets will have a back-to-back correlation in azimuthal angle $\Delta \phi$ which is extended in the rapidity over a plateau of $\Delta\eta \sim 2\ln\sqrt{s}/2p_T$. Higher order corrections to the LO pQCD results from initial state radiations and initial intrinsic transverse momentum of the two colliding parton will give rise to a broad back-to-back correlation of dijets in azimuthal angle $\Delta \phi$. Taking into account the transverse momentum from final state interaction and hadronization, one should expect to see a dihadron correlation that has a near-side two-dimensional peak for hadrons from the fragmentation of a single jet. Hadrons from separate fragmentation of two back-to-back jets  will give a back-side dihadron correlation that has a broad peak in azimuthal angle $\Delta \phi$ and an extended flat plateau in pseudo-rapidity $\Delta\eta$ in the shape of a ridge, therefore referred to as the ridge structure.

The above picture of dihadron correlations from minijets are based on the pQCD collinear parton model. However, recent experimental data at LHC show a ridge structure in the near-side dihadron correlation in the high multiplicity events of p+p collisions \cite{Khachatryan:2010gv} that cannot be explained within the above collinear parton model. Such a near-side ridge structure is also seen in high multiplicity events of p+Pb collisions at LHC \cite{CMS:2012qk,Abelev:2012cya,Aad:2012gla}. To explain such a surprising ridge structure in near-side dihadron correlation in p+p and p+A collisions, one can go beyond the collinear parton model and consider the initial transverse momentum distribution in a nucleon in the limit of gluon saturation at small momentum fraction $x$. The interference in multiple parton scattering is shown \cite{Dumitru:2010iy,Dusling:2012iga,Dusling:2012cg,Dusling:2012wy} to give rise to a near-side ridge in dihadron correlations that agrees well with the experimental measurements in p+p and p+Pb collisions at LHC. 

The charged hadron multiplicity in p+p collisions at $\sqrt{s}=7$ TeV has to be larger than 110 when one starts to see a ridge in the near-side dihadron correlation. This is equivalent to semi-central Cu+Cu collisions at the RHIC energy where collective expansion of a dense matter is clearly observed. Concentrated within a transverse area much smaller than in semi-central Cu+Cu collisions, the produced partons in the high multiplicity events of p+p collisions should experience final-state interaction that could lead to some form of collective behavior. Indeed, hydrodynamic models have been applied to p+p and p+A collisions and one can also qualitatively describe the measured dihadron correlations with a ridge-like long range correlation in rapidity \cite{Werner:2010ss,Bozek:2012gr,Qin:2013bha}. The most compelling evidence for collective flow in central p+Pb collisions at LHC is the observation of the mass (or flavor) dependence of the anisotropic flow in dihadron correlation \cite{Abelev:2012cya}. While the justifications for fast thermalization and the use of ideal hydrodynamic model for large multiplicity p+p and p+A collisions are still up to debate, it is safe not to consider such collective behavior or any initial state effect in minimum-biased events of p+p collisions in this section.

In this paper, we use associated hadron yields per-trigger-particle to study di-hadron correlations for charged hadrons in p+p and A+A collisions. This method is widely used in RHIC and LHC experiments and is described in details in Ref.~\cite{Chatrchyan:2011eka}. Trigger particles are defined as all charged hadrons within pseudo-rapidity window $|\eta|<2.4$ with transverse momentum $p_T^{\rm trig}$. The averaged number of trigger particles per event is denoted as $N_{\rm trig}$. Particle pairs are constructed by associating each trigger particle with all other charged hadrons referred to as associated particles within $|\eta|<2.4$ with transverse momentum $p_{T}^{\rm asso}$.

The dihadron correlation is defined as,
\begin{equation}
	C_{12}(\Delta\eta,\Delta\phi)= \frac{S(\Delta\eta, \Delta\phi)}{B(\Delta\eta, \Delta\phi)},
	\label{eq:npair}
\end{equation}
through ratios of associated yields per trigger, where $\Delta\eta, \Delta\phi$ are the differences in $\eta$ and $\phi$, respectively, of the trigger and associated particles. $S(\Delta\eta, \Delta\phi)$ is the signal associated yield per trigger particle in the same events,
\begin{equation}
	S(\Delta\eta, \Delta\phi) = \langle \frac{1}{N_{\rm trig}} \frac{d^{2}N^{\rm same}}{d\Delta\eta d\Delta\phi} \rangle,
	\label{eq:nsame}
\end{equation}
and $B(\Delta\eta, \Delta\phi)$ is the background associated yield per trigger-particle from mixed events,
\begin{equation}
	B(\Delta\eta, \Delta\phi) = \langle \frac{1}{N_{\rm trig}} \frac{d^{2}N^{\rm mixd}}{d\Delta\eta d\Delta\phi} \rangle,
	\label{eq:nmixd}
\end{equation}
where the hadron pair is constructed by associating trigger particles in one event with associated particles in another random event.
The average is carried out over events in which $N_{\rm trig}\neq 0$.

\begin{figure}[htp]
	\centering
	\includegraphics[width=0.45\textwidth]{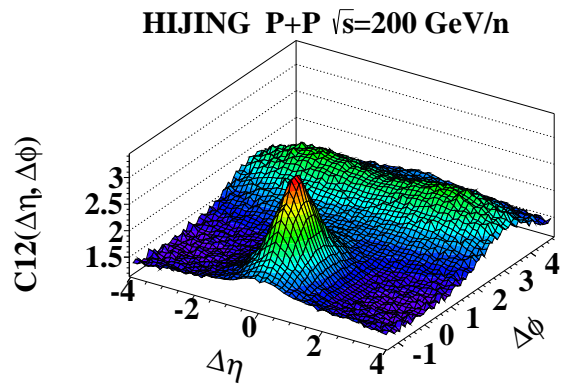}
	\caption{(Color online) Dihadron correlation for charged hadrons in p+p collisions at $\sqrt{s}=200$ GeV/n given by HIJING Monte Carlo model. The transverse momentum range for trigger and associated particles are $p_{T}^{\rm trig}\in (2,3) $ GeV/$c$ and $p_{T}^{\rm asso}\in (0,2) $ GeV/$c$, respectively.}
		\label{fig:pphijing1}
	\end{figure}

Shown in Fig.~\ref{fig:pphijing1} is the dihadron correlation for charged hadrons in p+p collisions at $\sqrt{s}=200$ GeV/n. 
The two-dimensional distribution has the basic structure of di-hadron correlation as expected from minijets.
There is one near-side peak from the single jet fragmentation and one away-side ridge at $\Delta\phi=\pi$ from back-to-back jets whose total longitudinal momentum is often nonzero in the center-of-mass frame of p+p collisions. Since HIJING employs the pQCD collinear parton model for jet production and does not allow any final-state interaction among produced partons in p+p collisions, there should not be any near-side ridge structure in dihadron correlation.
	
\begin{figure}[htp]
\centering
\includegraphics[width=0.45\textwidth]{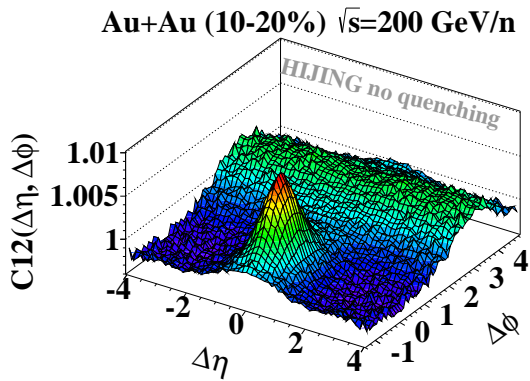}
\caption{(Color online) The same as Fig.~\ref{fig:pphijing1} except in semi-central (10-20\%) Au+Au collisions 
at $\sqrt{s}=200$ GeV/n from HIJING without jet quenching.}
\label{fig:AAhijing2}
\end{figure}

\begin{figure}[htp]
\centering
\includegraphics[width=0.45\textwidth]{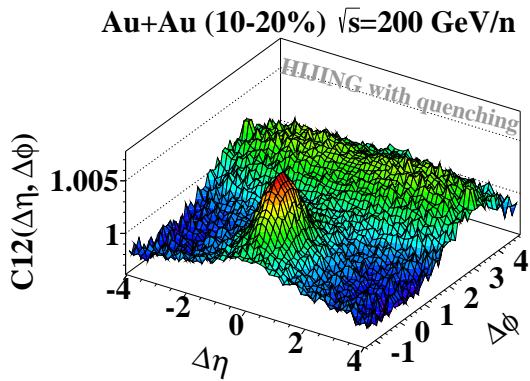}
\caption{(Color online) The same as Fig.~\ref{fig:pphijing1} except in semi-central (10-20\%) Au+Au collisions 
at $\sqrt{s}=200$ GeV/n from HIJING with jet quenching.}
\label{fig:AAhijing3}
\end{figure}

To study dihadron correlations from minijets in the environment of heavy-ion collisions without collective expansion, we show the HIJING results in semi-central (10-20\%) Au+Au collisions at $\sqrt{s}=200$ GeV/nucleon without and with jet quenching in Figs.~\ref{fig:AAhijing2} and \ref{fig:AAhijing3}, respectively. When jet quenching is turned off, there is no final-state interaction even in A+A collisions in HIJING. Minijet production is just superposition of binary nucleon-nucleon collisions. The nuclear modification of parton distributions in HIJNG does not affect dihadron correlations of the final state hadrons. The shape of the near-side peak in dihadron correlation in Au+Au remains the same as in p+p collisions. Because of the definition of the associated yield per trigger particle in Eq.~\ref{eq:npair}, the magnitude of dihadron correlation above the underlying background should inversely proportional to the total multiplicity. The ratio of back-side to near-side signal should on the other hand proportional to the average total number of minijets per event which is in turn also proportional to the number of binary nucleon-nucleon collisions in A+A collisions. One can see these two features through the comparison between Fig.~ \ref{fig:pphijing1} and \ref{fig:AAhijing2}.

\begin{figure}[htp]
\centering
\includegraphics[width=0.45\textwidth]{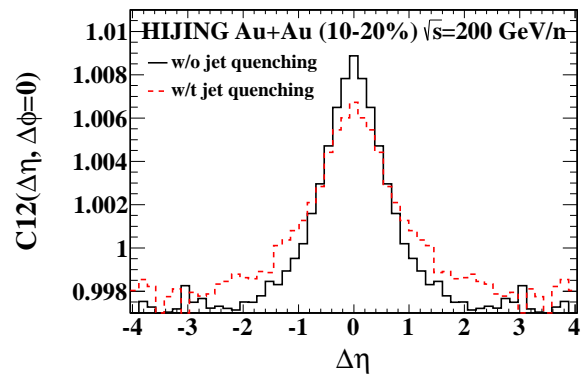}
\caption{(Color online) Near-side dihadron correlations for charged hadrons as a function of $\Delta\eta$ 
in central Au+Au collisions at $\sqrt{s}=200$ GeV/n in HIJING with (dashed) and without jet quenching (solid). The transverse momentum range for trigger and associated particles are $p_{T}^{\rm trig} \in (2,3)$ GeV/$c$ and $p_{T}^{\rm asso} \in (0,2) $ GeV/$c$, respectively.}
\label{fig:AAhijing4}
\end{figure}

\begin{figure}[htp]
\centering
\includegraphics[width=0.45\textwidth]{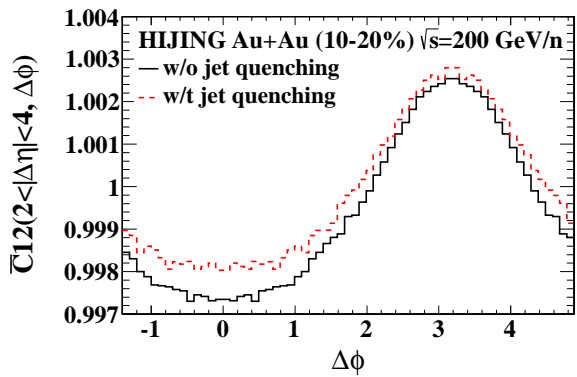}
\caption{(Color online) Dihadron correlations for charged hadrons with large pseudo-rapidity gap ($2<\Delta\eta<4$) as a function of $\Delta\phi$ in central Au+Au collisions at $\sqrt{s}=200$ GeV/n in HIJING with (dashed) and without jet quenching (solid). The transverse momentum range for trigger and associated particles are $p_{T}^{\rm trig} \in (2,3) $ GeV/$c$ and $p_{T}^{\rm asso} \in (0,2) $ GeV/$c$, respectively.} 
\label{fig:AAhijing5}
\end{figure}

Jet quenching in HIJING is implemented through a simple model of interaction between jet shower partons and the soft strings representing the bulk medium \cite{Wang:1991hta,Gyulassy:1994ew,Deng:2010mv,Xu:2012au}. It is expected that jet quenching in HIJING should diffuse parton distributions both inside jets and between jets. This will lead to diffusion of dihadron correlations. One can indeed see such diffusion in Fig.~\ref{fig:AAhijing3} for charged dihadron correlation in central Au+Au collisions with jet quenching as compared to that in Fig.~ \ref{fig:AAhijing2} without jet quenching. To illustrate quantitatively the diffusion of dihadron correlations due to jet quenching, we show in Fig.~\ref{fig:AAhijing4} the near-side peak ($\Delta\phi=0$) of the dihadron correlation as a function of $\Delta\eta$ in central Au+Au collisions from HIJING with (dashed) and without (solid) jet quenching.
The near-side peak along $\Delta\eta$ is clearly broadened, and the dihadron correlation outside the peak at $2<|\Delta\eta|<4$ is enhanced by jet quenching as compared to the case without jet quenching. To illustrate the enhancement of the dihadron correlation outside the jet-peak, we show in Fig.~\ref{fig:AAhijing5} the dihadron correlation with large rapidity gap ($2<\Delta\eta<4$). While the azimuthal angle distribution of the away-side ridge remains almost the same, there is clear enhancement of the dihadron correlation on the near-side due to jet quenching. However, this near-side enhancement is distributed evenly around $\Delta\phi=0$, therefore no ridge structure contrary to the momentum kick model \cite{Wong:2011qr}. Therefore, jet quenching cannot explain the long range correlation structure in p+p, p+A and A+A collisions at LHC \cite{Khachatryan:2010gv,CMS:2012qk,Abelev:2012cya,Aad:2012gla}. As we will show in the next section, the slight enhancement of near-side long range correlation due to jet quenching will be overwhelmed by ridge-like dihadron correlations from the collective flow of the expanding anisotropic medium. 

In addition to jet production, the underlying soft and coherent interaction also contributes to low transverse momentum hadron production in p+p collisions. Hadron production from this coherent process is modeled by fragmentation of strings between valence quarks and diquarks in HIJING. There should be a short range near-side dihadron correlation in rapidity for hadrons from such production mechanism. Momentum conservation also give rises to away-side dihadron correlation for these soft hadrons in p+p collisions. However, at high colliding energies such as RHIC and LHC, dihadron correlations from this soft and coherent process become insignificant, in particular at large transverse momentum, as compared to hadrons from minijets \cite{Wang:1992db}.

\section{Di-hadron correlations from (3+1)D event-by-event hydrodynamics}

Because of the large energy density produced in the early stage of heavy-ion collisions and strong interaction among the initially produced partons, one can assume approximate local thermalization during the early stage of heavy-ion collisions. The evolution of such anisotropic hot matter can then be described by hydrodynamic models. We have recently developed an ideal (3+1)D hydrodynamic model \cite{Pang:2012he} for heavy-ion collisions that uses fluctuating initial conditions from HIJING and parton cascade in AMPT model for event-by-event simulations.

\subsection{(3+1)D ideal hydrodynamic model}

Hydrodynamic models of high-energy heavy-ion collisions can be considered as effective models for the long wavelength dynamics of dense matter evolution. Local thermal equilibrium is assumed at some initial time $\tau_{0}$ and the evolution of the system afterwards can be described by conservation equations for energy-momentum tensor and net baryon current,
\begin{eqnarray}
\partial_{\mu}T^{\mu\nu}&=&0,\\
\partial_{\mu} J^{\mu}&=&0,
\end{eqnarray}
where the energy-momentum tensor and net baryon current can be expressed as
\begin{eqnarray}
T^{\mu\nu}&=&(\varepsilon+P)u^{\mu}u^{\nu}-Pg^{\mu\nu}, \nonumber \\ 
J^{\mu}&=&nu^{\mu},
\label{eq:hydro}
\end{eqnarray}
in terms of the local energy density $\varepsilon$, pressure $P$, the metric tensor $g^{\mu\nu}$, net baryon density $n$ (or any conserved charges) and time-like flow velocity $u^{\mu}$ with $u^2=1$. Short wavelength dynamics is included in the equation of state (EoS) for which
we will use the parameterization EoS s95p-v1 by Huovinen and Petreczky \cite{Huovinen:2009yb} based on lattice QCD calculations.

In high-energy heavy-ion collisions, minijets are a dominant source of the initial energy density that evolves into an expanding QGP.
Therefore, fluctuations in the number of minijets, their initial correlation and thermalization will dictate the later anisotropic expansion
of the fireball and the final dihadron correlations. To incorporate initial fluctuations and correlations from minijets in event-by-event (3+1)D hydrodynamic simulations, we will use the AMPT model \cite{Zhang:1999bd} to provide the local initial energy-momentum tensor in each hydrodynamic cell. The AMPT model uses the HIJING model \cite{Wang:1991hta,Gyulassy:1994ew,Deng:2010mv,Xu:2012au} to generate initial partons from hard and semi-hard scatterings and excited strings from soft interactions. 

We will use the $4$-momenta and spatial coordinates of partons from the AMPT model with a Gaussian smearing
function to determine the local energy-momentum tensor as the initial conditions for our event-by-event (3+1)D hydrodynamic simulations,
\begin{equation}
  \begin{aligned}
  T^{\mu\nu} &(\tau_{0},x,y,\eta_{s}) = K\sum_{i}
  \frac{p^{\mu}_{i}p^{\nu}_{i}}{p^{\tau}_{i}}\frac{1}{\tau_{0}\sqrt{2\pi\sigma_{\eta_{s}}^{2}}}\frac{1}{2\pi\sigma_{r}^{2}}\\
     	  &\hspace{-0.1in} \times \exp \left[-\frac{(x-x_{i})^{2}+(y-y_{i})^{2}}{2\sigma_{r}^{2}} - \frac{(\eta_{s}-\eta_{i s})^{2}}{2\sigma_{\eta_{s}}^{2}}\right],
  \end{aligned}
  \label{eq:Pmu}
\end{equation}
where $p^{\tau}_{i}=m_{iT}\cosh(Y_{i}-\eta_{i s})$, $p^{x}_{i}=p_{i x}$, $p^{y}_{i}=p_{i y}$ 
and $p^{\eta}_{i}=m_{i T}\sinh(Y_{i}-\eta_{i s})/\tau_{0}$ for parton $i$, which runs over all partons produced in the AMPT 
model simulations. 
We set $\sigma_{r}=0.6$ fm, $\sigma_{\eta_{s}}=0.6$ in our calculations. The transverse mass $m_{T}$, 
rapidity $Y$ and spatial rapidity $\eta_{s}$ are calculated from the parton's $4$-momenta and spatial coordinates.
The scale factor $K$ and the initial time $\tau_{0}$ are the only two parameters that are adjusted to fit the experimental data on central rapidity density of produced hadrons. The smearing is necessary to provide the initial condition from discrete particles to a continuous initial energy density distribution. It can also be considered as an effective process of local thermalization.

\begin{figure}[t]
\begin{center}
\includegraphics[scale=0.49]{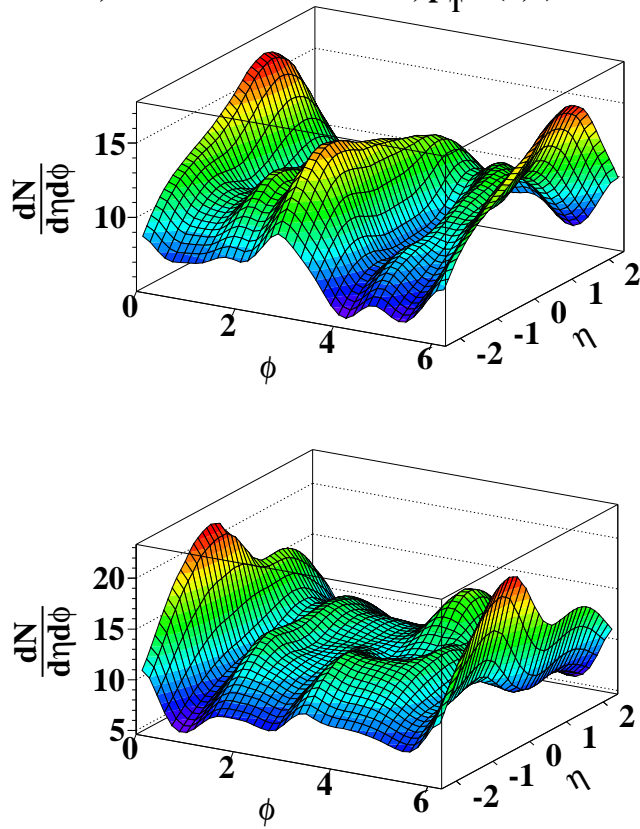}
\end{center}
\caption{(Color online) Charged hadron distributions in $\eta$ and $\phi$ within $p_{T} \in (1,2) $ GeV/$c$ from two typical hydrodynamic events of  $0-10\%$ Au+Au collisions at $\sqrt{s}=200$ GeV/n.}
\label{fig:dndetadphi}
\end{figure}

In the above initial condition from AMTP model, jets which consist of collimated clusters of partons in phase space will appear as a local density fluctuation or hot spots on top of the underlying background of bulk matter. Such local density fluctuations or hot spots will eventually contribute to the same-side dihadron correlations after anisotropic expansion according to the hydrodynamic equations. Since jets carry large transverse momentum (and longitudinal momentum in large rapidity region), these hot spots should also carry non-vanishing initial fluid velocity. Propagation of hot spots with non-vanishing fluid velocity will be equivalent to jet propagation through medium with the strongest jet-medium interaction and induced medium excitations. It should therefore influence the near-side dihadron correlation. Since jets are often produced in pairs, dijets in the above initial condition will appear as two hot spots with fluid velocities back-to-back in the transverse direction. Such back-to-back hot spots should contribute to away-side dihadron correlations in the final state.

Hadron spectra from the ideal hydrodynamics can be calculated through the Cooper-Frye formula \cite{Cooper:1974mv} at freeze-out for
particle $i$ with degeneracy $g_{i}$:
\begin{equation}
  \label{eq:frye-cooper}
  E\frac{dN_{i}}{d^{3}p}=\frac{dN_{i}}{d\eta p_{T}dp_{T}d\phi}=g_{i}\int_{\Sigma}p^{\mu}d\Sigma_{\mu}f_i(p\cdot u),
\end{equation}
where $d\Sigma_{\mu}$ is the normal vector on the freeze-out hyper-surface beyond which the temperature falls below the freeze-out temperature $T_{f}$ and hadrons are assumed to follow the thermal distribution,
\begin{equation}
  f_i(p\cdot u)=\frac{1}{(2\pi)^{3}}\frac{1}{e^{((p\cdot u - \mu_{i})/T_{f}))}\pm 1},
  \label{eq:fermion-bosen}
\end{equation}
where $\pm$ stands for fermions and bosons, respectively, $u$ is the flow velocity.
All resonances are assumed to freeze out from the same hyper surface and decay into stable particles whose spectra are summed together with direct hadrons from the freeze-out to give the final hadron spectra. Bulk hadron spectra and elliptic flow from this (3+1)D ideal hydrodynamic model are found to be in reasonable agreement with experimental data at RHIC and LHC \cite{Pang:2012he}.

\begin{widetext}
\begin{center}
\begin{figure}[h]
\includegraphics[scale=0.8]{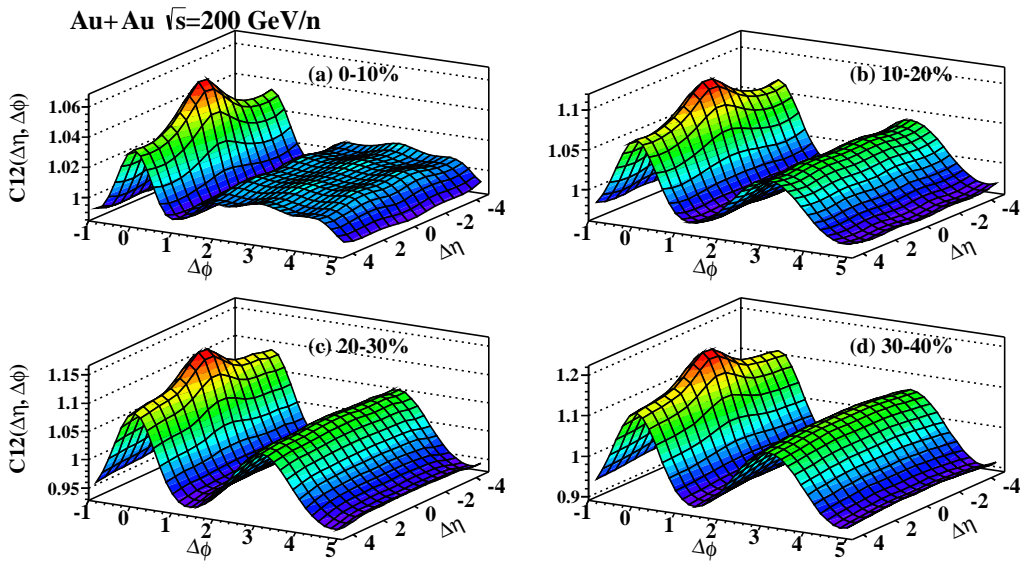}
\caption{(Color online) Dihadron correlations for charged hadrons from (3+1)D ideal hydrodynamic simulations of Au+Au collisions at $\sqrt{s}=200$ GeV/n with $4$ centralities. The trigger and associate particles lie in $p_T$ range $\in (2,4) $ GeV/$c$ and $\in (1,2)$ GeV/$c$, respectively.}
\label{fig:fc12}
\end{figure}
\end{center}
\end{widetext}

One can evaluate the above hadron spectra in each event with a given initial condition. The spectra should fluctuate in rapidity and azimuthal angle and vary from event to event with fluctuating initial conditions.  Shown in Fig.~\ref{fig:dndetadphi} are charged hadron distributions in $\eta-\phi$ within $p_T\in (1,2)$ GeV/$c$ from two typical hydrodynamic simulations of central $0-10\%$ Au+Au collisions at $\sqrt{s}=200$ GeV/n. These illustrate the typical azimuthal anisotropies of hadron production and their fluctuations in pseudo-rapidity $\eta$. The event-by-event fluctuations in azimuthal anisotropies and pseudo-rapidity distribution of charged hadron spectra from hydrodynamic simulations are completely dictated by the fluctuations in the initial conditions.  The smooth distributions do not contain statistical fluctuations due to finite number of particles within each bin.

One can calculate dihadron correlation in terms of bin-bin correlation in azimuthal and pseudo-rapidity, 
\begin{eqnarray}
  C_{12}(\Delta\eta,\Delta\phi) &=&  S(\Delta\eta, \Delta\phi) / B(\Delta\eta, \Delta\phi) \\
  &=& \frac{\langle N^{\rm trig}_{1}N^{\rm asso}_{2} / N^{\rm trig}_{\rm total}\rangle_{\rm same}}{\langle N^{\rm trig}_{1}N^{\rm asso}_{2} / N^{\rm trig}_{\rm total}\rangle_{\rm mix}} ,\label{eqn:c12}
  \label{eqn:C12_hydro}
\end{eqnarray}
where
\begin{equation}
  N^{\rm trig(asso)}_{i} = \int d p_T \frac{d^3 N^{\rm trig(asso)}_i} {d p_T d \eta_i d\phi_i}  d\eta_i d\phi_i, (i=1,2)
\end{equation}
denotes multiplicity for charged particles at pseudo-rapidity $\eta_{i}$ and azimuthal angle $\phi_{i}$ with transverse momentum integrated within the range for the trigger (associated) particles and $\langle \cdots \rangle_{\rm same(mix)}$ denotes averaging over events where trigger and associate particles come from the same event (different events). The above dihadron correlation is also averaged over all values of $\eta_{1,2}$ and $\phi_{1,2}$ with fixed difference $|\eta_{1}-\eta_{2}|=\Delta\eta$ and $|\phi_{1}-\phi_{2}|=\Delta\phi$ to increase the statistics and the final results should be symmetric in $\Delta\eta$ and $\Delta\phi$.
For mixed events, we randomly rotate each event in $\phi$ to remove the correlation caused by the same reaction plane used in the AMPT model. Comparing to the dihadron correlation in experiments and the HIJING calculation in Eq.~(\ref{eq:npair}), the above dihadron correlation from hydrodynamic simulations neglects statistical fluctuations due to finite number of hadrons within each bin. One may improve upon this calculation with statistic sampling within each bin in the calculation of the dihadron correlation. The statistical fluctuations might influence the overall magnitude of the dihadron correlation but not the shape. In addition, hadrons from resonance decays \cite{Bozek:2012en} also contribute to short-range correlations which we will neglect here. This can be taken into account with a Monte Carlo resonance decay within the statistic sampling in freeze-out. For moderately high $p_T$ ranges of trigger and associated hadrons that we consider in this study, the effect of resonance decay on dihadron correlations is expected to be small.

\subsection{Dihadron correlations at RHIC}

We first study dihadron correlations in heavy-ion collisions at RHIC.
Shown in Fig.~\ref{fig:fc12} are dihadron correlations for charged hadrons from event-by-event (3+1)D ideal hydrodynamic simulations of Au+Au at $\sqrt{s}=200$ GeV/n with initial conditions from AMPT model for $4$ different centralities. As we will show later, dihadron correlations shown in this figure are dominated by anisotropic collective flow due to the expansion of the fireball with fluctuating initial conditions. Since the dominant initial fluctuation of energy density in the transverse plane is geometric in nature, it should be common along the longitudinal direction in $\eta$ at each transverse position. Such coherent fluctuation from the initial parton production in HIJING \cite{Ma:2010dv} is the main cause of the ridge structures at both near-side and away-side of the dihadron correlation from event-by-event hydrodynamic simulations. As compared to HIJING results in the last section, such dihadron correlation generated by anisotropic flow is much stronger than the intrinsic dihadron correlation from minijets in heavy-ion collisions. Relics from minijets in these dihadron correlations from (3+1)D event-by-event hydrodynamic simulations appear as a peak at $\Delta\eta=0$ on top of the long ridge on the near-side.

For most non-central and peripheral collisions, elliptic flow $v_{2}$ is the most dominant and is driven by the overall geometric shape of the overlapping region of A+A collisions. This is the reason for the two ridge structures on both near-side and away-side of dihadron correlations. The amplitudes of these two ridges in $C_{12}$ increase with the average value of $v_{2}$ which in turn increases with the eccentricity as one can see in the figure. In the most central $0-10\%$ collisions, the amplitudes of $v_{3}$ and $v_{4}$ are comparable with $v_{2}$ for transverse momentum range $p_T\in (2,4)$ GeV/$c$ in the ideal (3+1)D hydrodynamic simulations \cite{Ma:2006fm,Ma:2010dv,Ma:2012zzh}. The large $v_{3}$ and $v_{4}$ component of the fluctuation is the reason for the double ridges on the away-side of the dihadron correlation as seen in the hydrodynamic results for the $0-10\%$ Au+Au collisions. 

\begin{figure}[h]
\includegraphics[scale=0.30]{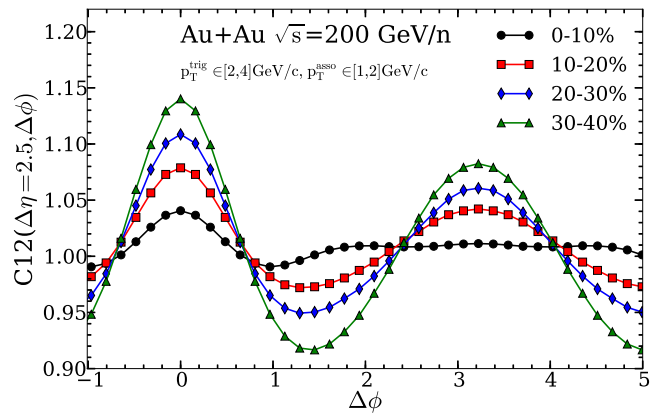}
\caption{(Color online) Dihadron correlations for charged hadrons with large rapidity gap $\Delta\eta=2.5$ as a function of $\Delta\phi$ from (3+1)D ideal hydrodynamic simulations of Au+Au collisions at $\sqrt{s}=200$ GeV/n with $4$ different centralities. The trigger and associate particles lie in $p_T$ range $\in (2,4)$ GeV/$c$ and $\in (1,2)$ GeV/$c$.}
\label{fig:c12slice}
\end{figure}

To illustrate dihadron correlations from anisotropic flows and their centrality dependence, we show in Fig.~\ref{fig:c12slice} dihadron correlations for charged hadrons with large rapidity gap $\Delta\eta=2.5$ as functions of $\Delta\phi$. The correlation strengths on both near-side and away-side increases from central to semi-peripheral events. For very peripheral events, viscous corrections become more important and the correlation should decrease again. In the most central collisions, higher harmonic flows become more dominant and the shape of away-side dihadron correlations becomes flatter with multiple bumps.

\begin{widetext}
\begin{center}
\begin{figure}[h]
  \includegraphics[scale=0.4]{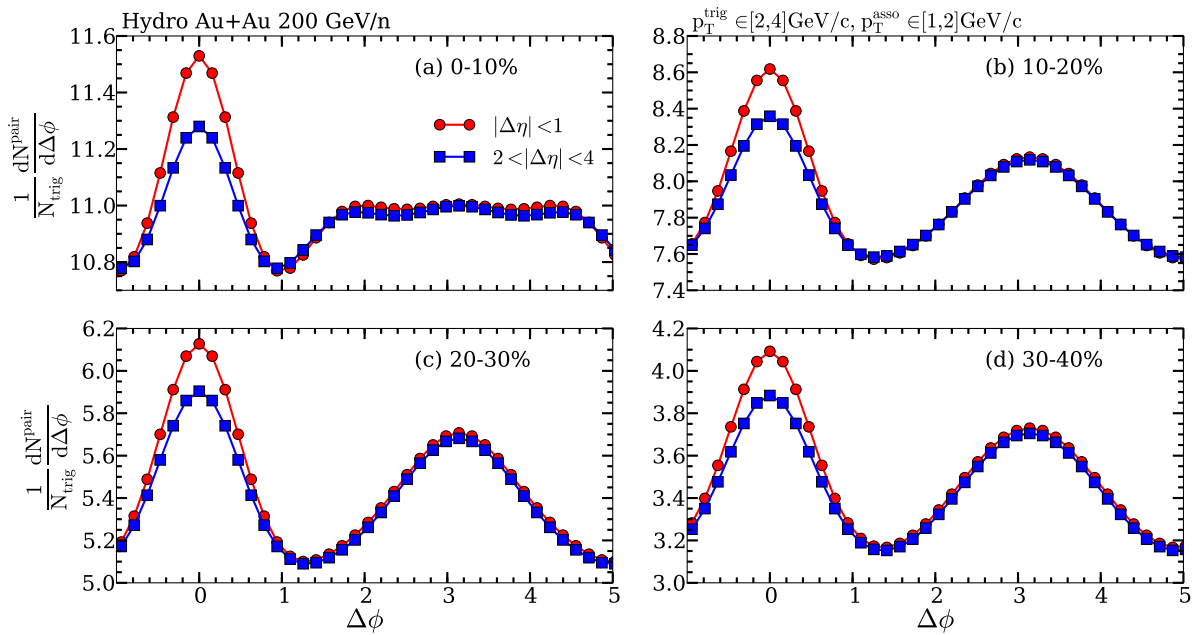}
\caption{(Color online) Associated yield per trigger for charged hadrons at short range $|\Delta\eta|<1$ (solid circles) and long range $2<|\Delta\eta|<4$ (solid squares) as a function of $\Delta\phi$ from (3+1)D ideal hydrodynamic simulations of Au+Au collisions at $\sqrt{s}=200$ GeV/n with $4$ different centralities. The trigger and associate particles lie in $p_T$ range $\in (2,4)$ GeV/$c$ and $\in (1,2)$ GeV/$c$, respectively.}
\label{fig:c12sliceB}
\end{figure}
\end{center}
\end{widetext}

To exam the structure of the minijet relics amid anisotropic flows, we define the associated yield per trigger and per unit
of phase space in ($\Delta\eta$,$\Delta\phi$),
\begin{equation}
  \frac{1}{N_{\rm trig}}\frac{d^2N^{\rm pair}}{d\Delta\eta d\Delta\phi} = C_{12}(\Delta\eta, \Delta\phi)\times \frac{B(0,0)}{\Delta\eta \Delta\phi},
  \label{eq:pertrigger}
\end{equation}
which is obtained by multiplying the dihadron correlation $C_{12}$ with the yield of background pairs from mixed event $B(\Delta\eta=0,\Delta\phi=0)$. Dividing by $\Delta \eta$ and $\Delta \phi$ will ensure the associate particle yield per trigger independent of the bin size. 
The per-trigger particle yield in one selected rapidity difference window as a function of azimuthal angle difference $\Delta\phi$ is defined as:

  \begin{equation}
  \frac{1}{N_{\rm trig}} \frac{dN^{\rm pair}}{d\Delta\phi}  = \frac{1}{\eta_{\rm max} - \eta_{\rm min}} \int_{\eta_{\rm min}}^{\eta_{\rm max}} \frac{1}{N_{\rm trig}}\frac{d^2N^{\rm pair}}{d\Delta\eta d\Delta\phi} d\Delta\eta
  \label{eq:npair_phi}
\end{equation}

In Fig.~\ref{fig:c12sliceB}, we plot the associated yields per trigger for charged hadrons as functions of the azimuthal angle difference $\Delta\phi$. We can see that the associated yields per trigger on near-side and away-side both increase with centrality monotonically, reflecting the increased collectivity in more central collisions.

Since minijets do not contribute to long-range dihadron correlations on the near-side as we have seen in p+p collisions without collective expansion, one can subtract the long-range (large $\Delta\eta$) correlation from the short-range (small $\Delta\eta$) to focus on the structure of dihadron correlation on the near-side purely from minijets. 
Shown in Fig.~\ref{fig:PHENIXcmp} are the long-range subtracted per trigger charged hadron yields as functions of $\Delta \phi$ from (3+1D) event-by-event hydrodynamic simulations of Au+Au collisions at $\sqrt{s}=200$ GeV/n (open circles) compared with PHENIX experimental data (solid diamonds)\cite{TakahitoTodorokiforthePHENIX:2013kia} and HIJING results on p+p collisions (dashed histograms). 
We consider the same kinematic cuts as in PHENIX analysis when we calculate 2-dimensional per trigger particle yields as functions of $\Delta\eta$ and $\Delta\phi$ for charged hadrons in $|\eta|<2.4$ from which the associated yield per trigger $(1/N_{\rm trig}) dN^{\rm pair}/d\Delta\phi$ as a function of $\Delta\phi$ is calculated following Eq.~(\ref{eq:npair_phi}). We then subtract the long-range associated yields per trigger in $2<|\Delta \eta|<4$ from the short-range ones in $|\Delta\eta|<1$. 
We see modest but systematic increase of long-range subtracted per trigger yield from peripheral to central heavy-ion collisions 
and significant enhancement over that in to p+p collisions. One might regard this as the consequence of interaction between minijets and the expanding medium as the radial flow boots the particle yields within the relics of minijets in hydrodynamic simulations.

\begin{widetext}
\begin{center}
\begin{figure}[h]
  \includegraphics[scale=0.4]{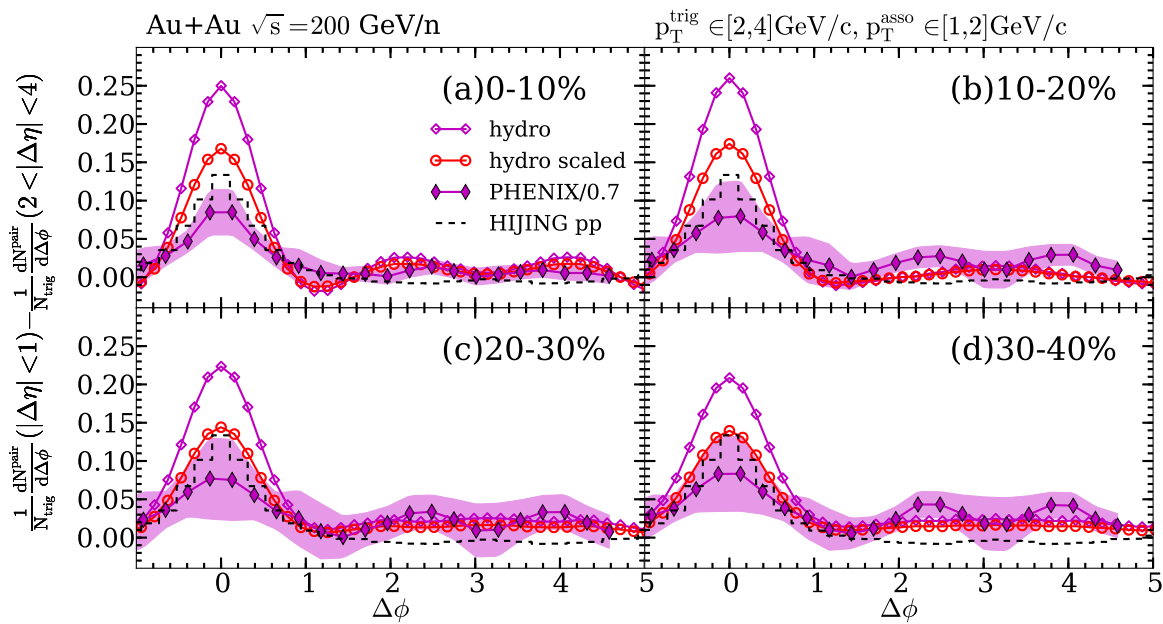}
\caption{(Color online) The short-range ($|\Delta\eta|<1$ ) yield per trigger with long-range ($2<|\Delta\eta|<4$) subtracted for charged hadrons as a function of $\Delta\phi$ from (3+1)D ideal hydrodynamic simulations of Au+Au collisions at $\sqrt{s}=200$ GeV/n with $4$ centralities (open diamonds and open circles), compared with HIJING p+p results (dashed histogram) and PHENIX data \cite{TakahitoTodorokiforthePHENIX:2013kia} (solid diamonds) where $v_2, v_3, v_4(\Psi_4)$ contributions are ZYAM subtracted from the short range per-trigger particle yield. The trigger and associated particles lie in $p_T$ range $\in (2,4)$ GeV/$c$ and $\in (1,2)$ GeV/$c$, respectively. See text for explanation on scaled hydro results.}
\label{fig:PHENIXcmp}
\end{figure}
\end{center}
\end{widetext}

\begin{center}
\begin{figure}[h]
 \includegraphics[width=0.5\textwidth]{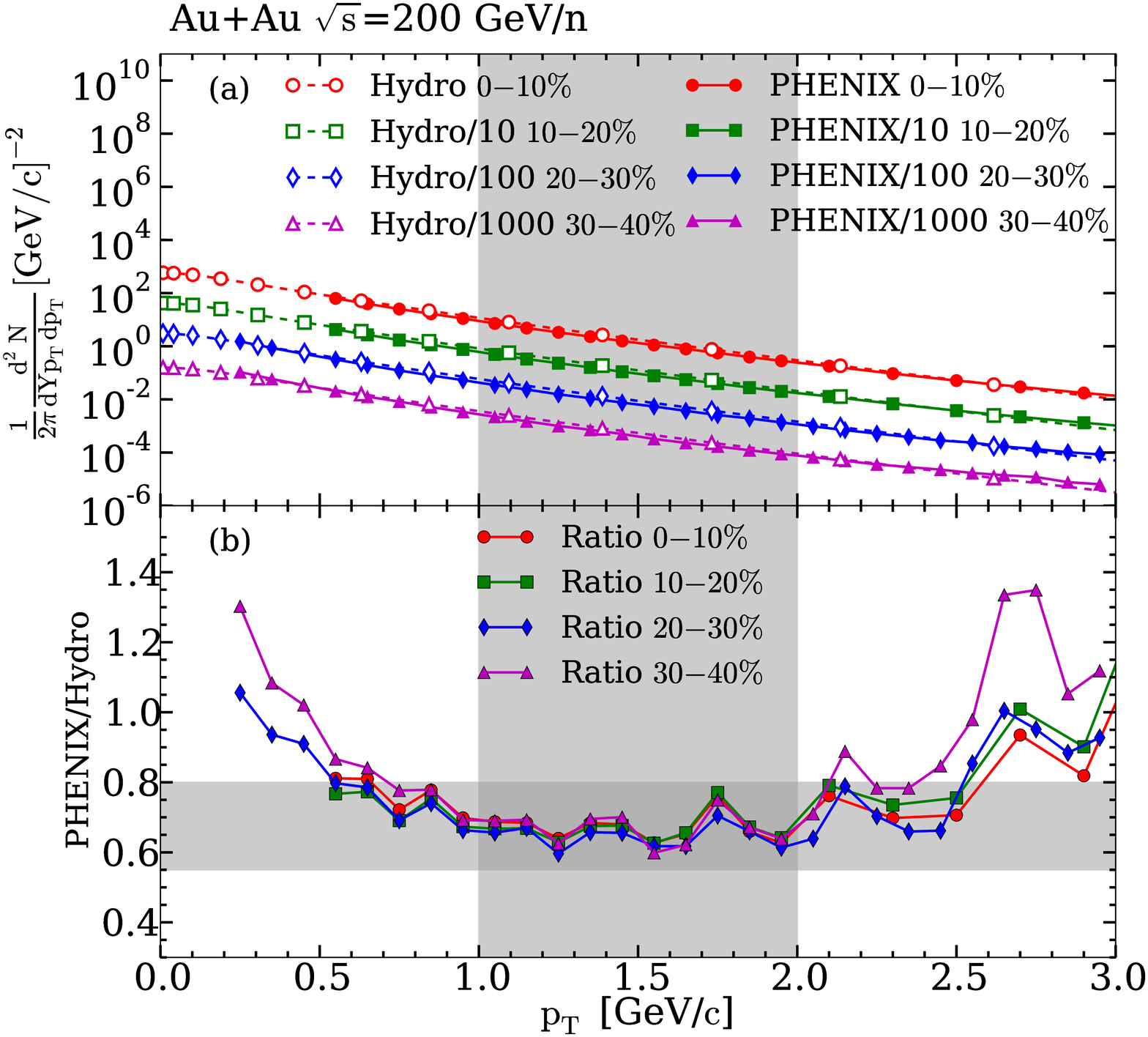}
 \caption{(Color online) (upper panel) The transverse momentum distribution of the charged hadrons in Au+Au collisions at $\sqrt{s}=200$ GeV/n from event-by-event (3+1)D hydrodynamic simulations (open symbols) \cite{Pang:2012he} compared with PHENIX experimental data (solid symbols) \cite{Adler:2003cb,Adare:2013esx} and (lower panel) their ratio. The shaded region indicates the $p_T$ range for associated hadrons in our calculation of dihadron correlations}
 \label{fig:ratio_phenix_hydro}
\end{figure}
\end{center}

In the PHENIX experiment, charged hadrons within $|\eta|<0.35$ are used to calculate the di-hadron correlations as functions of $\Delta \phi$, where the flow contributions $v_2, v_3, v_4(\Psi_4)$ are subtracted by using the ZYAM(Zero Yield At Minimum) method \cite{Adler:2005ee}. The data (solid diamonds) shown in Fig.~\ref{fig:PHENIXcmp} are the ZYAM subtracted per trigger particle yield and the hashed region represents systematic uncertainties propagated from higher-order flow harmonics. The PHENIX data shown in Fig.~\ref{fig:PHENIXcmp} are divided by a factor of $0.7$ due to the small rapidity window $|\eta|<0.35$ in order to get the associated yield per unit of rapidity. The long-range subtracted per trigger yields from event-by-event hydrodynamic simulations in Fig.~\ref{fig:PHENIXcmp} are systematically higher than the PHENIX results. The difference could be caused by the ZYAM method of subtracting background from high order harmonic flows.

The magnitude of the per-trigger particle yield is determined by the event averaged number of associated particles which in turn
is proportional to single inclusive hadron spectra in hydrodynamic simulations. We compare our ideal hydrodynamic simulations \cite{Pang:2012he} of $p_T$ spectra for charged hadrons in Au+Au collisions at $\sqrt{s}=200$ GeV/n to PHENIX data in Fig.~\ref{fig:ratio_phenix_hydro}. In the region $p_T \in (1,2)$ GeV/$c$, hydrodynamic results are consistently larger than experimental data. Inclusion of viscous corrections in hydrodynamic simulations will likely improve the hydrodynamic results on $p_T$ spectra. For a more accurate comparison to experimental data on associated yield per trigger, we should take into account such overestimate of hadron spectra in ideal hydrodynamic simulations. For this purpose we define a scale factor $C$ as the ratio between the integrated number of associated hadrons from experiments and event-by-event ideal hydrodynamic simulations in the range of transverse momentum $\Delta p_T$ of interest,
\begin{equation}
  C = \frac{(\int^{\Delta p_T} dp_T d^2N/d\eta dp_T)_{\rm Expt}}{(\int^{\Delta p_T} dp_T d^2N/d\eta dp_T )_{\rm Hydro}}.
  \label{eq:sfactor}
\end{equation}
In Table.~\ref{tab:sfactor_rhic}, we list these scale factors for Au+Au collisions at $\sqrt{s}=200$ GeV/n for $4$ centralities.
 The scaled hydrodynamic results on associated yield per trigger are plotted in Fig.~\ref{fig:PHENIXcmp} which are on the average about 30\% below the original hydrodynamic results.

\begin{table}[h]
  \centering
  \begin{tabular}{|c|c|c|c|c|}
    \hline
  Centrality &  $0-10\%$  &   $10-20\%$   &  $20-30\%$  & $30-40\%$  \\
    \hline
   C & 0.671 & 0.670 & 0.645 & 0.669  \\
    \hline
  \end{tabular}
  \caption{The scale factor $C$ which is defined as the ratio between the integrated number of charged hadrons with $p_T \in (1,2)$ GeV/$c$ from PHENIX data and event-by-event ideal hydrodynamic simulations in Fig.~\ref{fig:ratio_phenix_hydro}.}
  \label{tab:sfactor_rhic}
\end{table}

Long-range subtracted hadron yields per trigger have also been measured in Au+Au and d+Au collisions by STAR experiment at 
RHIC \cite{Agakishiev:2010ur,Agakishiev:2011st} and found to be similar. This might not be surprising given recent discovery of collective behavior such as anisotropic flows and ridge structures in p+Pb collisions at LHC \cite{CMS:2012qk,Abelev:2012cya,Aad:2012gla,Abelev:2012cya} and d+Au collisions at RHIC \cite{Adare:2013piz}. It is therefore important to compare results in A+A and p(d)+A to p+p collisions.

We should note that the subtraction of the long-range correlation removes contributions from all order harmonic flows to di-hadron correlation on both near side ($\Delta\phi=0$) and away side ($\Delta\phi=\pi$) as well as the jet contribution on the 
away-side. Since uncertainties from high order flow harmonics exist at both short-range $|\Delta\eta|<1$ and long-range $2<|\Delta\eta|<4$, the long-range subtraction method should significantly reduce the systematic errors arising from direct and high harmonic flows as compared to the ZYAM method where only flow contributions are subtracted. This is particularly important for our calculations in this paper since ideal hydrodynamic models are known to produce larger direct and high order harmonic flows \cite{Schenke:2011bn} than experimental data. Inclusion of viscosity will improve the hydrodynamic calculation of high harmonic flows and it might also influence quantitatively the long-range subtracted correlations. Long-range subtracted dihadron correlations can also avoid uncertainties related to the ZYAM method for subtraction of flow contributions \cite{Luzum:2010sp}.

\subsection{Dihadron correlations at LHC}

For Pb+Pb collisions at the LHC energy $\sqrt{s} = 2.760$ TeV/n, we also calculate the per trigger charged yield as a function of $\Delta\eta$ and $\Delta\phi$. At such a high colliding energy, the initial energy density is much higher than at RHIC. There are also much more mini-jets contributing to the fluctuation and correlation in the initial conditions for hydrodynamic studies.


\begin{widetext}
\begin{center}
\begin{figure}[h]
\includegraphics[scale=0.8]{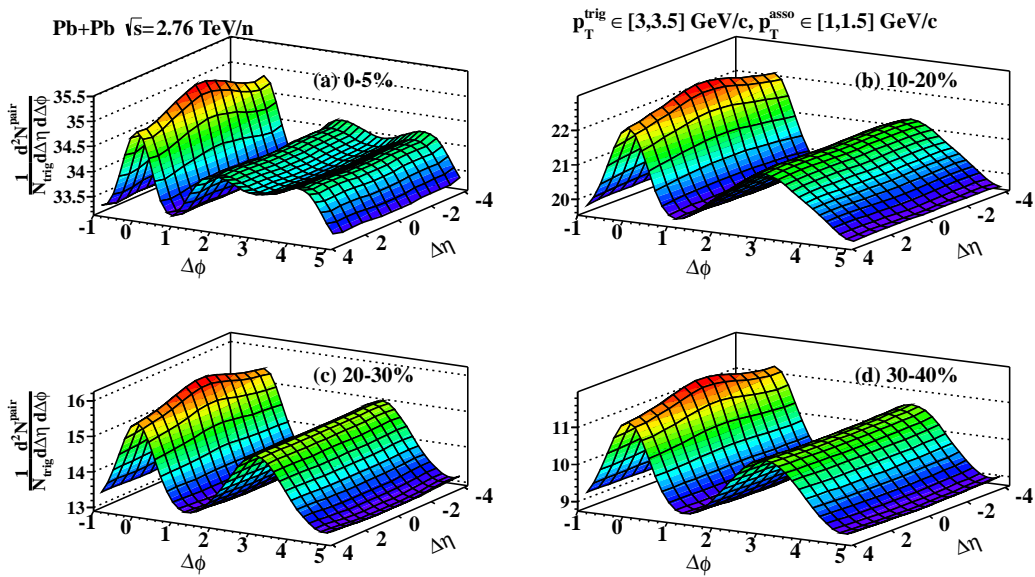}
\caption{(Color online) Associated yield of charged hadrons per trigger from (3+1)D ideal hydrodynamic simulations of Pb+Pb collisions at $\sqrt{s}=2.76$ TeV/n with $4$ different centralities. The trigger and associate particles lie in $p_T$ range $\in (3,3.5)$ GeV/$c$ and $ \in (1,1.5)$ GeV/$c$, respectively.}
\label{fig:c12PbPb}
\end{figure}
\end{center}
\end{widetext}

\begin{center}
\begin{figure}[h]
 \includegraphics[width=0.5\textwidth]{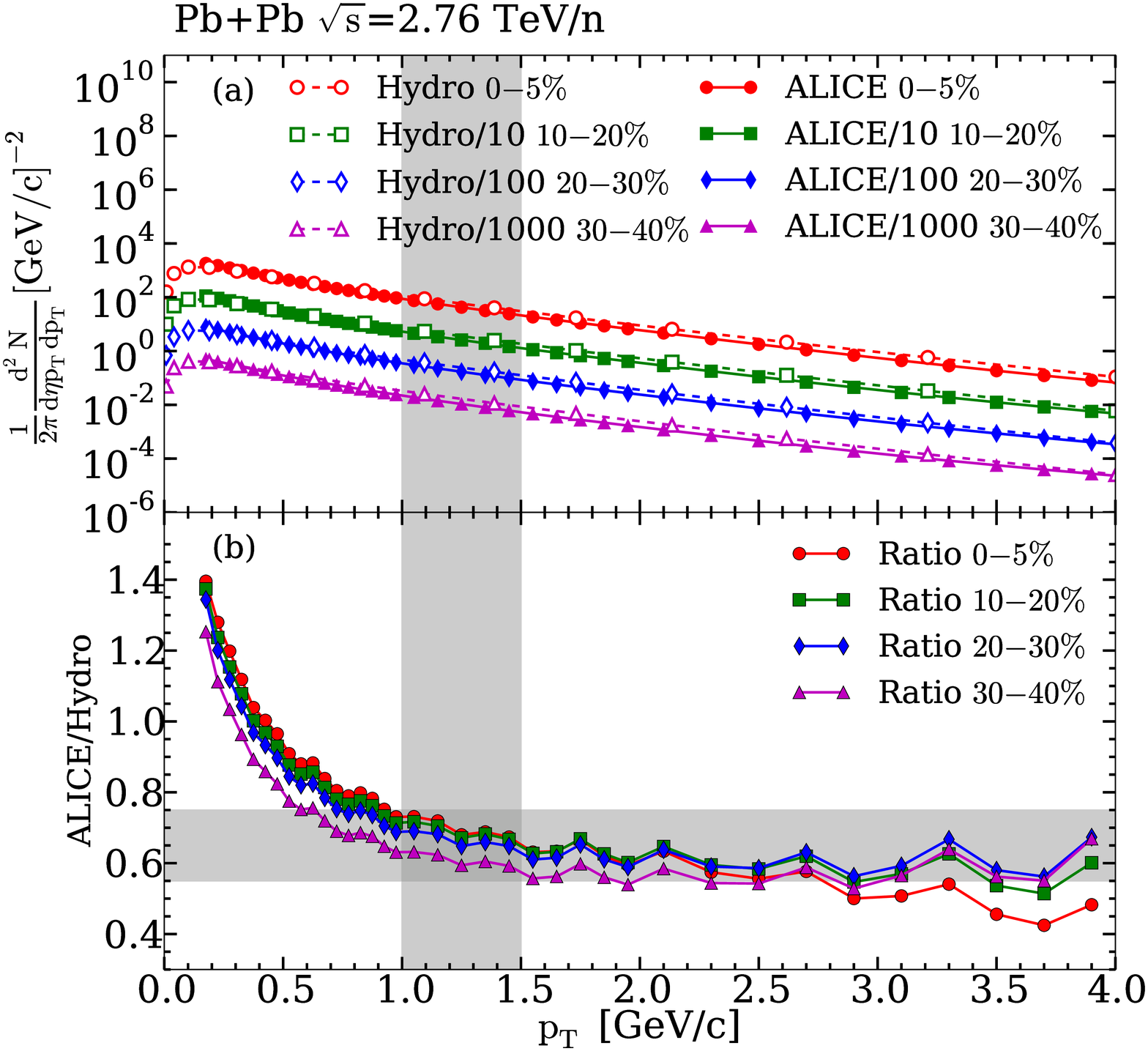}
 \caption{(Color online) The transverse momentum distribution of the charged hadrons in Pb+Pb collisions at $\sqrt{s}=2.76$ TeV/n from event-by-event (3+1)D hydrodynamic simulations (open symbols) \cite{Pang:2012he} compared with ALICE experimental data (solid symbols) \cite{Abelev:2012hxa}. The shaded region indicates the $p_T$ range for associated hadrons in our calculation of dihadron correlations.}
 \label{fig:ratio_alice_hydro}
\end{figure}
\end{center}

Shown in Fig.~\ref{fig:c12PbPb} are the per trigger particle yield in Pb+Pb collisions at $\sqrt{s}=2.76$ TeV/n for $4$ centralities from event-by-event (3+1D) ideal hydrodynamic simulations, where the transverse momentum ranges for trigger and associate particles are $(3,3.5)$ GeV/$c$ and $(1,1.5)$ GeV/$c$, respectively.
Again,  slopes of $p_T$ spectra for charged hadrons from our ideal hydrodynamic simulations \cite{Pang:2012he} of Pb+Pb collisions at $\sqrt{s}=2.76$ TeV/n are a little bigger than experimental data as shown in Fig.~\ref{fig:ratio_alice_hydro}. They also make the event-averaged number of associate particles in the transverse momentum range $(1,1.5)$ GeV/$c$ from hydrodynamic simulations bigger than experimental data. This brings about $25\%$ more overall per-trigger particle yield in the most central collisions and $35\%$ more in semi-central collisions than CMS data \cite{Chatrchyan:2012wg}. Viscous corrections in hydrodynamic simulations will bring down the $p_T$ spectra and give a better fit to the experimental data on the magnitudes of per trigger particle yield, especially in peripheral collisions. 

To compensate the effect of viscous corrections to the overall spectra and the magnitude of per trigger hadron yields, we scale our per trigger associated particle yields from ideal hydrodynamic simulations by a scale factor $C$ for each centrality range in Eq.~(\ref{eq:sfactor}).  It is defined as the ratio between the integrated number of associated hadrons from ALICE \cite{Abelev:2012hxa} and event-by-event ideal hydrodynamic simulations \cite{Pang:2012he}. In Table.~\ref{tab:sfactor_lhc}, we list these scale factors for Pb+Pb collisions at $\sqrt{s}=2.76$ TeV/n for $4$ centralities. 

\begin{table}[h]
  \centering
  \begin{tabular}{|c|c|c|c|c|}
    \hline
  Centrality & $0-5\%$  &   $10-20\%$   &  $20-30\%$  & $30-40\%$  \\
    \hline
   C & 0.728 & 0.717 & 0.693  & 0.635  \\
    \hline
  \end{tabular}
  \caption{The scale factor $C$ defined as the ratio between the integrated number of charged hadrons with $p_T \in (1,1.5)$ GeV/$c$ from ALICE data and event-by-event ideal hydrodynamic simulations in Fig.~\ref{fig:ratio_alice_hydro}.}
  \label{tab:sfactor_lhc}
\end{table}

\begin{widetext}
\begin{center}
\begin{figure}[h]
  \includegraphics[scale=0.4]{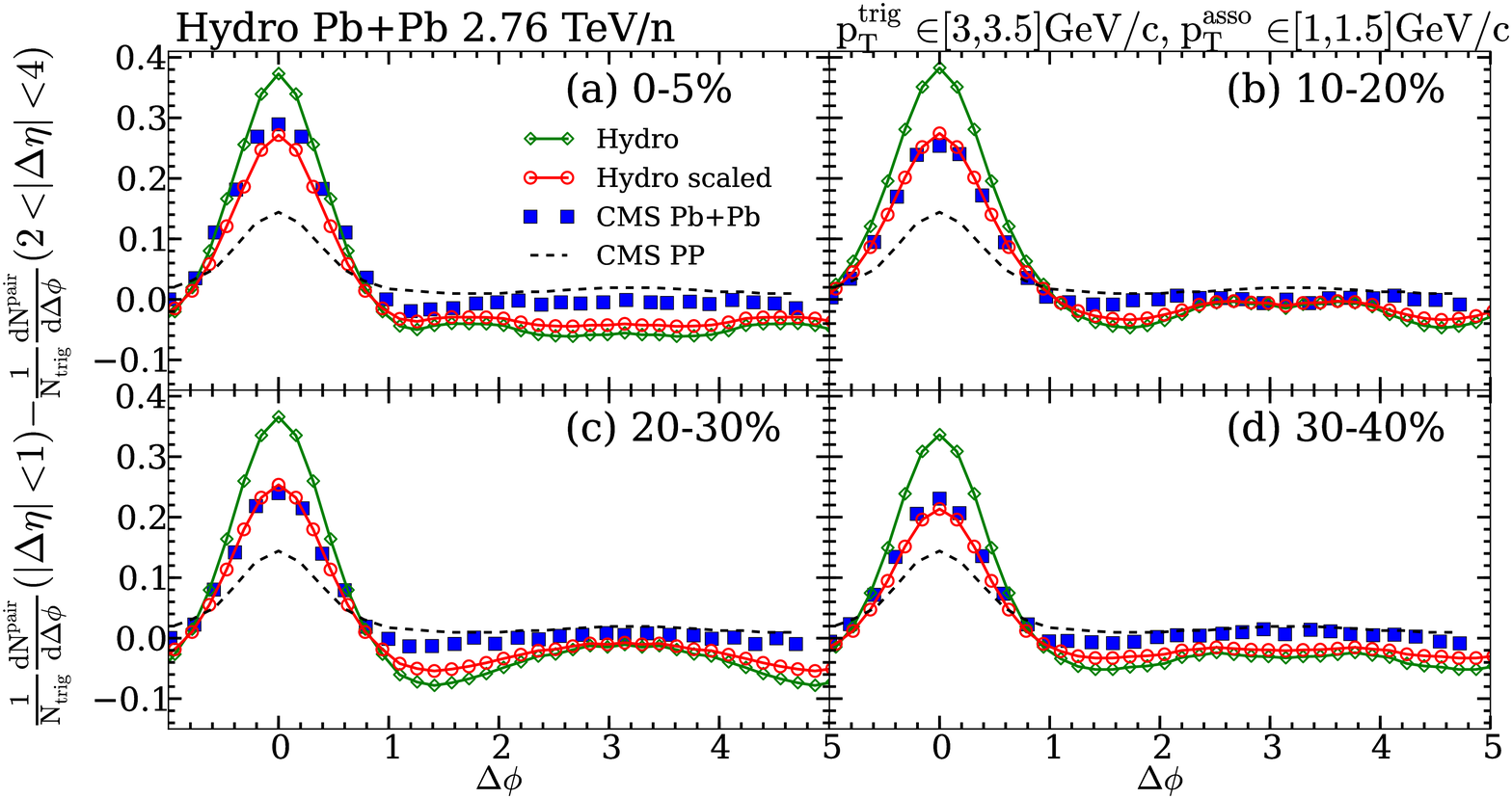}
\caption{(Color online) The difference between associated yield per trigger for charged hadrons in $|\Delta\eta|<1$ and $2<|\Delta\eta|<4$ as a function of $\Delta\phi$ from ideal hydrodynamic simulations (open circles and open diamonds) of Pb+Pb collisions at $\sqrt{s}=2.76$ TeV/n with $4$ centralities, compared with CMS data for Pb+Pb (solid squares) and p+p (dashed) collisions \cite{Chatrchyan:2012wg}. The trigger and associated particles lie in $p_T$ range $\in (3,3.5)$ GeV/$c$ and $\in (1,1.5)$ GeV/$c$, respectively. See text for explanations on scaled hydro results.}
\label{fig:c12_cmp_CMS}
\end{figure}
\end{center}
\end{widetext}

In Fig.~\ref{fig:c12PbPb}, we again see the difference between short-range and long-range associated hadron yield per trigger on the near-side $|\Delta\phi|<1.1$ due to relics of minijets. On the away side at $\Delta\phi\sim \pi$, the ridge of per trigger particle yield is approximately flat from mostly the superposition of anisotropic flows and some small contributions of back-to-back jets. However,
a careful examination reveals that there is a small concave feature along the ridge on the away-side, especially for most central collisions, which is also observed in CMS \cite{Chatrchyan:2012wg} data for Pb+Pb collisions at $2.76$ TeV/n and STAR \cite{Agakishiev:2011pe} data for Au+Au collisions at $200$ GeV/n. A smaller but similar concave feature along $\Delta \eta$ on the away side is also seen in HIJING results for p+p collisions at 200 GeV in Fig.~\ref{fig:pphijing1}. We have also checked that such a concave feature along the away-side ridge still exists if true rapidity $y$ is used instead of pseudo-rapidity $\eta$. One possible mechanism for the enhanced concave feature along the away-side ridge in heavy-ion collisions is the influence of longitudinal expansion on back-to-back jet correlations. This might worth further investigation in future studies.

To study the centrality dependence of minijet relics, we show in Fig.~\ref{fig:c12_cmp_CMS} the long-range subtracted associated hadron yields per trigger from Fig.~\ref{fig:c12PbPb} (open diamonds) compared with CMS experimental data for Pb+Pb (solid squares) and p+p (dashed lines) at $2.76$ TeV/n. The difference between hydrodynamic results and experimental data on the magnitudes of the near-side yield per trigger is consistent with the difference in the overall single hadron spectra in the $p_T$ range of the associated hadrons, as we have noted before. Using the scale factors in Table.~\ref{tab:sfactor_lhc}, the scaled hydrodynamic results (open circles) agree with the CMS data very well on the long-range subtracted per-trigger hadron yields. Compared to the p+p result (dashed lines), the per trigger hadron yield from minijet relics in heavy-ion collisions is significantly higher. One can attribute such enhanced per trigger yields to the jet-medium interaction and the influence of radial flows on minijet relics. The enhancement in Pb+Pb collisions decreases in more peripheral collisions and eventual should approach that in p+p collisions. We should note again that the subtraction of long-range dihadron correlations reduces the influence of higher harmonic flows on the near-side correlation and consequently of viscous corrections to the ideal hydrodynamic results.


\section{Relics of minijets and longitudinal fluctuations}

As we have seen in the last section, relics of minijets produce additional dihadron correlations that sit on top of two ridges from anisotropic flows due to expansion of the fluctuating initial energy density that is coherent along the longitudinal direction.
Such relics of minijets in dihadron correlations result from the initial correlation intrinsic to minijets that survive anisotropic expansion of the fluctuating fireball. In this section, we will exam the structure of minijet relics in detail.

\subsection{Longitudinal fluctuations and initial flow}

 In order to study the influence of the intrinsic correlations on final dihadron correlations, we first switch off the longitudinal fluctuation and correlation from minijets by multiplying the initial parton density at central rapidity $\eta=0$ with an envelope distribution function along $\eta$ direction,
\begin{equation}
H(\eta)= \exp\left[-\theta(|\eta|-\eta_{0})(|\eta|-\eta_{0})^2/2\sigma_{w}^{2}\right],
\label{eq:tube}
\end{equation}
to obtain a tube-like initial rapidity distribution, where $\eta_{0}$ is the half width of the central plateau in rapidity and $\sigma_{w}$ 
is the Gaussian fall-off at large rapidity that are used to fit the charged hadron rapidity distribution \cite{Pang:2012he}.

\begin{figure}[t]
\begin{center}
\includegraphics[scale=0.6]{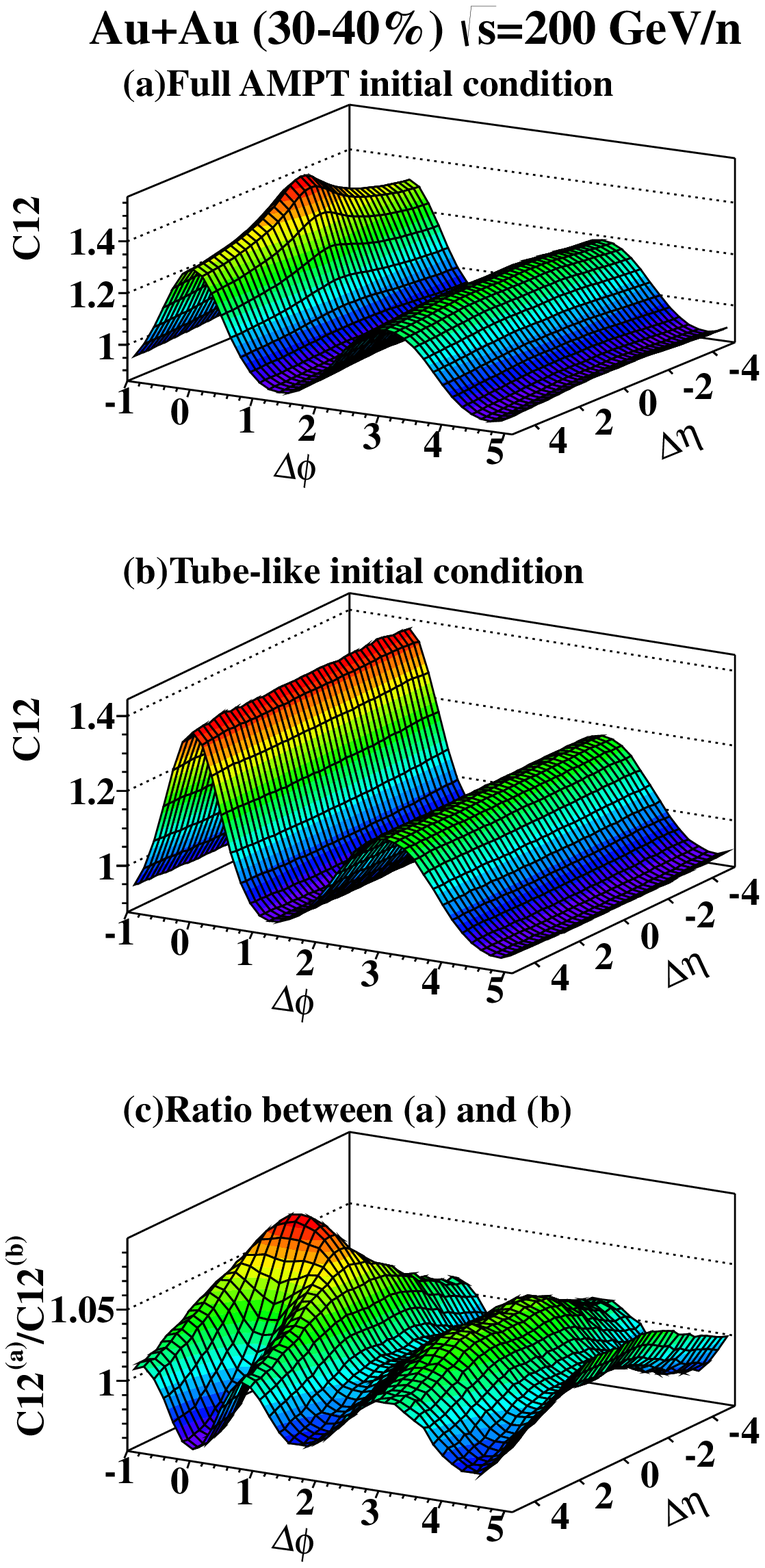}
\end{center}
\caption{(Color online) Dihadron correlations from event-by-event ideal hydrodynamic simulations with full AMPT initial conditions (upper) and tube-like AMPT initial conditions (middle) in semi-central $30-40\%$ Au+Au collisions at $\sqrt{s}=200$ GeV/n, and their ratio (lower). The trigger and associated particles lie in the same momentum range $p_{T} \in (2,3) $ GeV/$c$. }
\label{fig:c12_tube}
\end{figure}

In Fig.~\ref{fig:c12_tube} we compare dihadron correlations from event-by-event (3+1)D hydrodynamic simulations with the full AMPT fluctuating initial conditions (upper panel) and tube-like AMPT initial conditions (middle panel) in semi-central $30-40\%$ Au+Au collisions at $\sqrt{s}=200$ GeV/n. Within the rapidity window of the plot, the tube-like AMPT initial conditions give two ridges which are very flat along $\Delta\eta$ direction. Therefore, any variation in rapidity of the dihadron correlation as compared to that from the tube-like initial conditions are caused by initial intrinsic correlation from minijets and the underlying soft and coherence initial parton production. To illustrate the relics of minijets in dihadron correlation, we show in the lower panel of Fig.~\ref{fig:c12_tube} the ratio of dihadron correlation from hydrodynamic simulations with full AMPT and tube-like AMPT initial conditions. The ratio has a two-dimensional peak on the near-side and long-ridge on the away-side. They resemble the dihadron correlation from minijets in p+p collisions shown in Fig.~\ref{fig:pphijing1}, though the near-side peak of the ratio is much broader in $\eta$ direction than that  in p+p collisions. Such broadening of the near-side peak in the relic dihadron correlation from minijets represents the effect of thermalization and jet-medium interaction. Mechanisms such as local charge conservation in hadronization and resonance decay \cite{Bozek:2012en} can also induce two-dimensional central peak in dihadron correlations which is additional to the relics of minijets.

\begin{figure}[t]
\begin{center}
\includegraphics[scale=0.6]{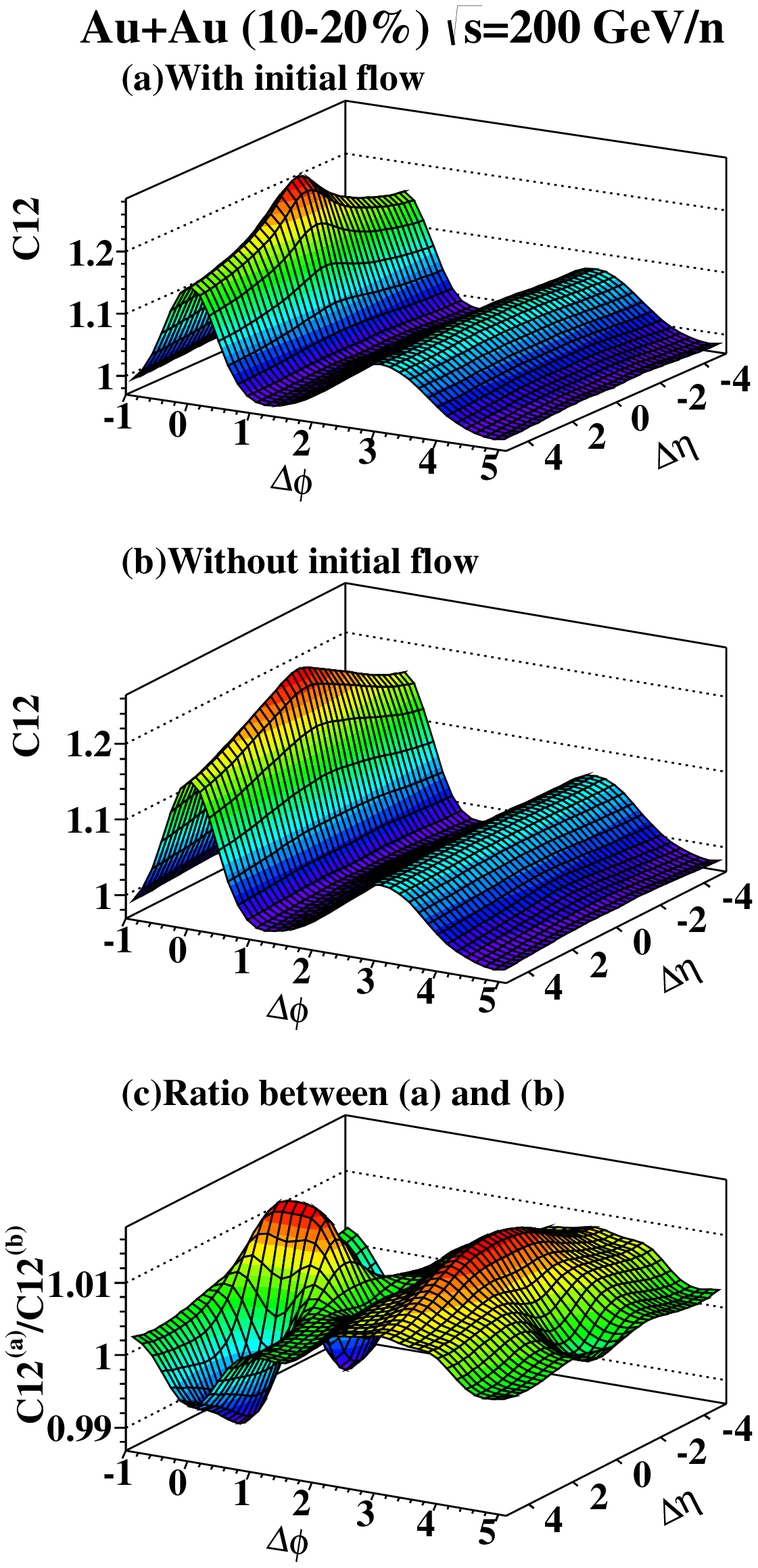}
\end{center}
\caption{(Color online) Dihadron correlations in central $10-20\%$ Au+Au collisions at $\sqrt{s}=200$ GeV/n from ideal hydrodynamic simulations with full AMPT initial conditions (upper) and AMPT initial conditions without initial transverse fluid velocity (middle) and their ratio (lower). The trigger and associated particles lie in the same momentum range $p_{T} \in (2,3) $ GeV/$c$.}
\label{fig:c12_no_iniflow}
\end{figure}

Because of momentum conservation, transverse momentum of jets will be transferred to the local medium through jet-medium interaction even if one assumes a complete local thermalization. Therefore, hot spots from minijets in the fluctuating initial condition should have non-vanishing fluid velocity. This non-vanishing fluid velocity of hot spots is found to increase the final hadron multiplicity, the slope of hadron transverse momentum spectra and the differential elliptic flow at large transverse momentum \cite{Pang:2012he}. It should also affect both near-side and away-side dihadron correlations. Shown in Fig.~\ref{fig:c12_no_iniflow} are dihadron correlations with full fluctuating initial condition from AMPT (upper panel) and initial conditions in which the local transverse fluid velocity is set to zero (middle panel). One can see that the near-side peak in the case of full fluctuation is more collimated in pseudo-rapidity $\Delta\eta$ than that without initial flow. To quantify the effect of initial flow in the dihadron correlation we show the ratio of dihadron correlations with and without initial flow in the lower panel of Fig.~\ref{fig:c12_no_iniflow}. One can clearly see the relic of minijets in this ratio whose structure resembles that of dihadron correlations in Au+Au collisions from HIJING simulations in Fig.~\ref{fig:AAhijing4} which do not have contributions from anisotropic flow. However, the effect of the initial flow on the dihadron correlation is quite small on the order of a few percent of the overall magnitude of dihadron correlations.

\subsection{Dihadron correlations and harmonic flows}

Though our analysis of (3+1)D hydrodynamic simulations with different initial conditions can illustrate the relics of minijets in dihadron correlations on both near side and away side, the striking feature of dihadron correlations that is unambiguously from minijets and can be extracted from experimental data is the near-side peak on top of the ridges from harmonic flows. An alternative method to quantify the difference in dihadron correlations within and outside the near-side peak region is to carry out a harmonic analysis of dihadron correlations.

We use the event plane as constructed from hadrons at large rapidity $3.3<|\eta|<4.8$ in Pb+Pb collisions at the LHC energy $\sqrt{s}=2.76$ TeV/n,
\begin{equation}
\Psi_{n}^{\rm EP}= \frac{1}{n}\arctan \frac{\langle p_{T}\sin(n\phi) \rangle}{\langle p_{T}\cos(n\phi) \rangle},
\label{eq:eplane} 
\end{equation}
where the average,
\begin{equation}
  \langle {\cal O}(p_T,\eta) \rangle=\frac{\int_{0}^{2\pi}\,d\phi\;  {\cal O}(p_T,\eta,\phi) \frac{dN}{d\eta p_{T}dp_{T}d\phi}}{\int_{0}^{2\pi}\,d\phi \frac{dN}{d\eta p_{T}dp_{T}d\phi}},
  \label{eq:vnaverage}
\end{equation}
for each event is over the azimuthal angle of all final hadrons weighted by their spectra. Note that final hadron spectra from the Coorper-Frye formula are continuous distribution functions. Therefore, the calculation of $\Psi_{n}^{\rm EP}$ in our hydrodynamic simulations will not introduce plane resolution, contrary to experimental analyses where there are only finite number of particles per event. Harmonic flows with respect to the event plane are defined as
\begin{equation}
  v_{n}^{\rm EP}(p_{T},\eta)=\langle\!\langle \cos\bigl(n(\phi-\Psi_{n}^{\rm EP})\bigr)\rangle\!\rangle,
  \label{eq:vn_pt}
\end{equation}
where additional average over events is implied. In our current event-by-event (3+1)D ideal hydrodynamic simulations, higher order harmonic flows
are always larger than experimental data. Introduction of viscosity in a viscous hydrodynamics will bring down higher harmonic flows \cite{Schenke:2011bn}. For our purpose of study here, dihadron correlations can be reconstructed from harmonic flows within the same hydrodynamic events self-consistently.

With harmonic flows determined from hydrodynamic events, one can construct the corresponding dihadron correlation as,
\begin{eqnarray*}
  C_{12}^{\rm EP}(\Delta\phi) &=&  a_0 \cos(\Delta\phi) + b_0 \left[1 + 2 \sum_{n=2}^{6}v_{n}^{\rm EP}v_{n}^{\rm EP}\cos(n\Delta\phi)\right] \\
  C_{12}^{22}(\Delta\phi) &=&  a_1 \cos(\Delta\phi) + b_1 \left[1 + 2 \sum_{n=2}^{6}v_{n}^{\rm 22}v_{n}^{\rm 22}\cos(n\Delta\phi)\right].
\label{eq:corr_vn}
\end{eqnarray*}
where $C_{12}^{\rm EP}(\Delta\phi)$ and $C_{12}^{22}(\Delta\phi)$ are constructed from event-plane harmonic flows $v_{n}^{\rm EP}$ and mean-square-root harmonic flows $v_{n}^{22}$, respectively. 
The mean-square-root harmonics flows are defined as $v_{n}^{22} = \sqrt{\langle v_{n}^{\rm EP}v_{n}^{\rm EP} \rangle}$ and take the event-by-event fluctuations into account.
Since we can not determine the directed flow $v_{1}$ in our hydrodynamic calculations we will just adjust parameters $a_0, a_1$ and $b_0, b_1$ to fit to the calculated raw dihardron correlations. The fitting parameters are shown in Table.~\ref{tab:c12_dec}.

\begin{table}[h]
  \centering
  \begin{tabular}{|c|cc|cc|}
    \hline
  $\ $  &  $a_0$  &   $b_0$   &  $a_1$  & $b_1$  \\
    \hline
0-5\%   &  0.00754 & 1.0098 & 0.00754  & 1.00967  \\
    \hline
10-20\% &  0.01049 & 1.00597 & 0.01049 & 1.00524  \\
    \hline
  \end{tabular}
  \caption{The fitting parameters $a_0, b_0$ and $a_1, b_1$ for harmonic flow decompositions of dihadron correlations in Pb+Pb collisions at $\sqrt{s}=2.76$ TeV/n with (0-5\%) and (10-20\%) centrality.}
  \label{tab:c12_dec}
\end{table}

\begin{figure}[t]
\begin{center}
\includegraphics[scale=0.6]{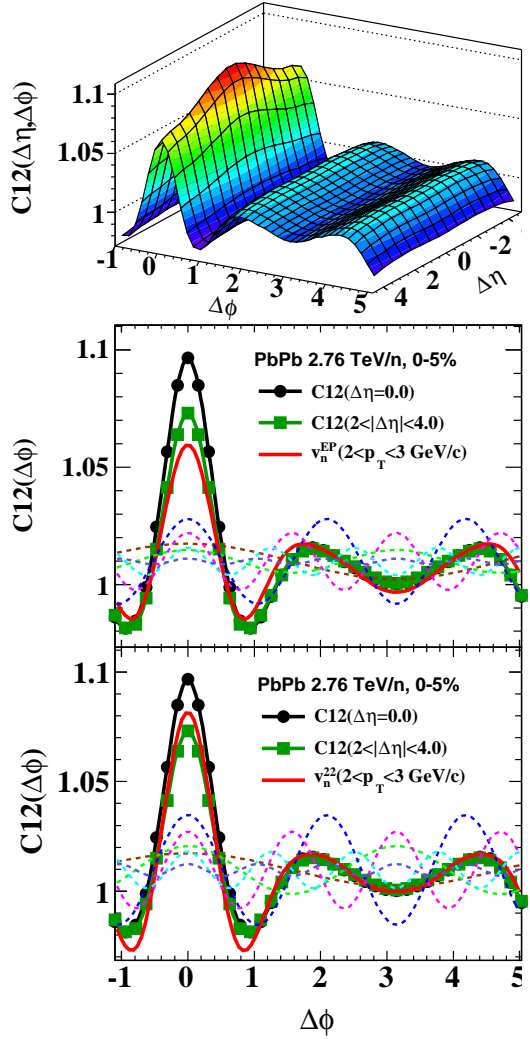}
\end{center}
\caption{(Color online) The $v_n$ reconstruction of dihadron correlations for $0-5\%$ central Pb+Pb collisions at $\sqrt{s}=$2.76 TeV/n with $ p_T \in (2,3)$ GeV/$c$.}
\label{fig:c12_cent_vn1}
\end{figure}

\begin{figure}[t]
\begin{center}
\includegraphics[scale=0.6]{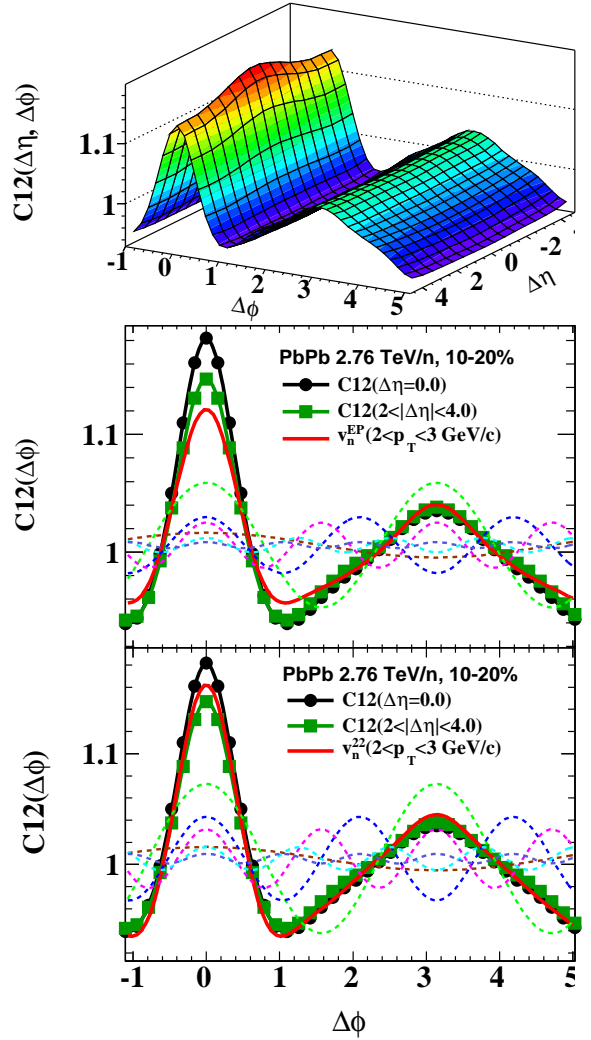}
\end{center}
\caption{(Color online) The same as Fig~\ref{fig:c12_cent_vn1} except for seminal central $10-20\%$ Pb+Pb collisions.}
\label{fig:c12_cent_vn2}
\end{figure}

Shown in Figs.~\ref{fig:c12_cent_vn1} and \ref{fig:c12_cent_vn2} (middle and bottom panels) are directly calculated dihadron correlations 
with zero rapidity gap (solid circles ) and large rapidity gap $2<|\Delta\eta|<4$ (solid squares) as compared to dihadron 
correlations reconstructed from harmonic flows (solid lines) in central (0-5\%) and semi-central (10-20\%) Pb+Pb 
collisions at $\sqrt{s}=2.76$ TeV/n. The momentum range for both trigger and associated hadrons is $p_{T} \in (2,3)$ GeV/$c$. The corresponding 2-D dihadron correlations are shown in the top panels for illustrations. Harmonic flows $v_{n}^{\rm EP}$ and $v_{n}^{\rm 22}$ for $n=2-6$ are also calculated with $p_T$ integrated over the same range. Contributions from each harmonic flow to dihadron correlations are superimposed in the figure (dashed lines). 

As we can see from Fig.~\ref{fig:c12_cent_vn1} and \ref{fig:c12_cent_vn2}, there is very little contribution from minijets to dihadron correlations with large rapidity gap $2<|\Delta\eta|<4$ where harmonic flows from anisotropic expansion dominate. 
Therefore, raw dihadron correlations (solid-square-lines) can be described well by correlations reconstructed from mean-square-root harmonic flows (solid lines in the bottom panels) in both central and semi-central collisions. 
Short-range ($\Delta\eta=0$) dihadron correlations (solid-circle-lines) on the away-side also agree very well with that reconstructed from harmonic flows. But they differ significantly at near-side, indicating strong non-flow effect in short-range dihadron correlations. The excess in the raw dihadron correlations over contributions from harmonic flows is from the relics of minijets in heavy-ion collisions. 

In central collisions (Fig.~\ref{fig:c12_cent_vn1}), triangle flow ($n=3$) is the most dominant contribution to dihadron correlations while the quadratic flow is comparable to elliptic flow at the selected transverse momentum range. The reconstructed dihadron correlation from harmonic flows ($n=2-6$) therefore has a double bump on the away-side, corresponding to the double ridges on the away-side in the two-dimensional dihadron correlation. In semi-central and peripheral collisions (Fig.~\ref{fig:c12_cent_vn2}), however, elliptic flow becomes dominant as a result of the overall geometric shape of the overlapping region. Dihadron correlations on the away-side have a single peak, corresponding to a single away-side ridge in the two-dimensional dihadron correlations.

\section{Conclusions and Discussions}

We have used both the HIJNG Monte Carlo model and an ideal (3+1)D hydrodynamic model with fluctuating initial conditions from HIJING+AMPT model to calculate two-dimensional dihadron correlations in $\Delta\eta$ and $\Delta\phi$. We investigated the influence of initial local fluctuations from minijets on the final dihadron correlations.
Since HIJING does not have final-state interaction to produce any collective expansion of the bulk medium, dihadron correlations in A+A from HIJING are found to have a similar structure as those in p+p collisions. It has a peak on the near side and a ridge along $\Delta\eta$ on the away side. Jet quenching is found to broaden the near-side peak in both $\Delta\eta$ and $\Delta\phi$ but does not produce any ridge structure on the near side. 

Within a (3+1)D ideal hydrodynamic model with fluctuating initial conditions from the HIJING+AMPT model, we found that dihadron correlations are dominated by harmonic flows from the expansion of the anisotropic fireball, especially for charged hadrons with large rapidity gap where influence from minijets is expected to be small. Short-range dihadron correlations with small rapidity gap show similar two-ridge structure, however, with an enhanced dihadron correlation on the near side due to intrinsic correlations from minijets in the fluctuating initial conditions. These intrinsic correlations seem to survive the anisotropic expansion and still show up in near-side dihadron correlations in final states. The intrinsic away-side correlation from minijets, however, seems all disappear in the final state after hydrodynamic evolution. This is consistent with the picture of jet quenching which suppresses the back-to-back dihadron correlation since the away-side jets have to traverse a large volume of dense matter, while the near-side correlation remains about the same because of trigger bias toward surface emission of jets with little attenuation. 

Since near-side dihadron correlations from minijets are limited to short range in rapidity, one can extract their contributions by subtracting long-range correlations due to anisotropic flows. The near-side correlations from relics of minijets extracted with this method are not sensitive to values of harmonic flows and therefore their sensitivity to viscous corrections should also be reduced. The long-range subtracted per trigger hadron yields on the near side are found to be significantly enhanced in central heavy-ion collisions over that in p+p, due to influence of radial flow during the hydrodynamic evolution. The hydrodynamic results on the long-range subtracted per trigger hadron yields are in qualitative agreement with experimental data at RHIC and LHC. 

By comparing dihadron correlations from hydrodynamic simulations with full AMPT initial conditions, tube-like AMPT initial conditions and AMPT initial conditions without initial fluid velocity, we illustrated the influence of minijets on the final dihadron correlations on both near side and away side. The longitudinal correlations and fluctuations in minijets have a much stronger effect on the near-side dihadron correlation than the away-side. The effect of initial fluid velocity from back-to-back dijets enhances the dihadron correlation about 1\% on both near-side and away-side. We also used harmonic flows calculated from the same hydrodynamic events to reconstruct dihadron correlations and found they can describe the raw long-range (with large rapidity gap) dihadron correlations very well, whereas they differ from the raw short-range dihadron correlations because of contributions from relics of minijets, as expected.

We should emphasize that our study of dihadron correlations within the (3+1)D ideal hydrodynamic model is still semi-quantitative. It has been shown \cite{Schenke:2011bn} that viscous corrections to the harmonic flows and therefore dihadron correlations are large in particular for higher harmonic flows at high transverse momentum. Therefore, our ideal hydrodynamic results on the overall dihadron correlations and per trigger hadron yields will differ from experimental data by the amount of viscous corrections, especially in higher $p_T$ regions.
We should emphasize that our study is only limited to low transverse momentum region. At much higher $p_T$, non-equilibrium corrections become too big and one has to resort to other approaches such as transport or jet quenching models.

\section*{Acknowledgement}

We would like to thank helpful discussions with Wei Li, Victor Roy and Fuqiang Wang.
This work was supported by the National Natural Science Foundation of China
under the grant No. 11221504 and 11125524, and by the Director, Office of Energy
Research, Office of High Energy and Nuclear Physics, Division of Nuclear Physics, of the U.S. Department of 
Energy under Contract No. DE-AC02-05CH11231 and within the framework of the JET Collaboration,

\end{document}